%% file: paper-ieee.tex
\documentclass[10pt, conference, letterpaper]{IEEEtran}
\usepackage{tikz}
\usepackage{truncate} 
\usepackage{filecontents}
\usepackage[most]{tcolorbox}

\usepackage[colorlinks, citecolor=blue, hyperfootnotes=false]{hyperref}
\usepackage{setspace, amsmath, url, lscape, algorithmic, multirow, pslatex, listings, verbatim, alltt, amsfonts, wrapfig, boxedminipage, color, bookmark,comment}
\usepackage{xcolor} 
\usepackage[vlined,linesnumbered,ruled,boxed]{algorithm2e}
\usepackage{array}
\usepackage{subcaption}

\DeclareFontFamily{\encodingdefault}{\ttdefault}{\hyphenchar\font=`\-}
\newcommand\ProcessThreeDashes{\llap{\color{cyan}\mdseries-{-}-}}

\usepackage{paralist}
\usepackage{epsfig}
\let\labelindent\relax
\usepackage[inline]{enumitem}

\usepackage{xcolor}

\usepackage{listings}

\newcommand\YAMLcolonstyle{\color{red}\mdseries}
\newcommand\YAMLkeystyle{\color{black}\bfseries}
\newcommand\YAMLvaluestyle{\color{blue}\mdseries}

\makeatletter

\newcommand\language@yaml{yaml}

\expandafter\expandafter\expandafter\lstdefinelanguage
\expandafter{\language@yaml}
{
  keywords={true,false,null,y,n},
  keywordstyle=\color{darkgray}\bfseries,
  basicstyle=\ttfamily,                                 
  sensitive=false,
  comment=[l]{\#},
  morecomment=[s]{/*}{*/},
  commentstyle=\color{purple}\ttfamily,
  stringstyle=\YAMLvaluestyle\ttfamily,
  moredelim=[l][\color{orange}]{\&},
  moredelim=[l][\color{magenta}]{*},
  moredelim=**[il][\YAMLcolonstyle{:}\YAMLvaluestyle]{:},   
  morestring=[b]',
  morestring=[b]",
  literate =    {---}{{\ProcessThreeDashes}}3
                {>}{{\textcolor{red}\textgreater}}1     
                {|}{{\textcolor{red}\textbar}}1 
                {\ -\ }{{\mdseries\ -\ }}3,
}

\lst@AddToHook{EveryLine}{\ifx\lst@language\language@yaml\YAMLkeystyle\fi}
\makeatother

\DeclareMathDelimiter{(}{\mathopen} {operators}{"28}{largesymbols}{"00}
\DeclareMathDelimiter{)}{\mathclose}{operators}{"29}{largesymbols}{"01}

\newcommand{\qedd}{\nobreak \ifvmode \relax \else
      \ifdim\lastskip<1.5em \hskip-\lastskip
     \hskip1.5em plus0em minus0.5em \fi \nobreak
      \vrule height0.75em width0.5em depth0.25em\fi}

\newcommand{\comments}[1]{}
\newcommand\hl{\bgroup\markoverwith
  {\textcolor{yellow}{\rule[-.5ex]{2pt}{2.5ex}}}\ULon}

\begin{document}


\title{Benchmarking Message Brokers for IoT Edge Computing: A Comprehensive Performance Study}

\author{
\IEEEauthorblockN{Tapajit Chandra Paul\textsuperscript{1}, Pawissanutt Lertpongrujikorn\textsuperscript{1}, Hai Duc Nguyen\textsuperscript{2}, and Mohsen Amini Salehi\textsuperscript{1}}
\IEEEauthorblockA{\textsuperscript{1}{High Performance Cloud Computing \href{https://hpcclab.org/}{(HPCC)} Lab}, University of North Texas, USA\\
\textsuperscript{2}Argonne National Laboratory and University of Chicago, USA\\
\{tapajitchandra.paul, pawissanutt.lertpongrujikorn, mohsen.aminisalehi\}@unt.edu, hai.nguyen@anl.gov}
}





\maketitle
\IEEEpeerreviewmaketitle

\input{Sources/abstract}


\input{Sources/introduction}
 \input{Sources/background}
\input{Sources/related-work}
\input{Sources/experimental-design}

\input{Sources/evaluation}
\input{Sources/conclusion}
\input{Sources/acknowledgement}

%


\bibliographystyle{IEEEtran}
\bibliography{references}


\end{document}

%% file: Sources/abstract.tex
\begin{abstract}
Asynchronous messaging is a cornerstone of modern distributed systems, enabling decoupled communication for scalable and resilient applications. Today’s message queue (MQ) ecosystem spans a wide range of designs, from high-throughput streaming platforms to lightweight protocols tailored for edge and IoT environments. Despite this diversity, choosing an appropriate MQ system remains difficult.
Existing evaluations largely focus on throughput and latency on fixed hardware, while overlooking CPU and memory footprint and the effects of resource constraints, factors that are critical for edge and IoT deployments. In this paper, we present a systematic performance study of eight prominent message brokers: Mosquitto, EMQX, HiveMQ, RabbitMQ, ActiveMQ Artemis, NATS Server, Redis (Pub/Sub), and Zenoh Router. We introduce \textbf{mq-bench}, a unified benchmarking framework to evaluate these systems under identical conditions, scaling up to 10,000 concurrent client pairs across three VM configurations representative of edge hardware. This study reveals several interesting and sometimes counter-intuitive insights. Lightweight native brokers achieve sub-millisecond latency, while feature-rich enterprise platforms incur 2--3$\times$ higher overhead. Under high connection loads, multi-threaded brokers like NATS and Zenoh scale efficiently, whereas the widely-deployed Mosquitto saturates earlier due to its single-threaded architecture. We also find that Java-based brokers consume significantly more memory than native implementations, which has important implications for memory-constrained edge deployments. Based on these findings, we provide practical deployment guidelines that map workload requirements and resource constraints to appropriate broker choices for telemetry, streaming analytics, and IoT use cases.
\end{abstract}

\begin{IEEEkeywords}
Message Brokers, Publish/Subscribe, Edge Computing, Internet of Things (IoT), Benchmarking, Performance Evaluation, Distributed Systems, MQTT.
\end{IEEEkeywords}


%% file: Sources/introduction.tex
\section{Introduction}\label{sec:intro}
The proliferation of Internet of Things (IoT) devices, edge computing platforms, and microservice architectures has fundamentally transformed how modern applications communicate. From smart factories monitoring thousands of sensors to connected vehicles exchanging real-time telemetry, from healthcare wearables streaming patient vitals~\cite{wu2019wearable} to smart city infrastructure coordinating traffic flow, the demand for reliable, low-latency messaging infrastructure continues to grow exponentially. Industry analysts project tens of billions of connected devices by 2030~\cite{iot_analytics_2024}, each generating data that must be collected, processed, and acted upon in near real-time. At the heart of these systems lies asynchronous messaging---a paradigm that enables loosely coupled components to exchange data without blocking, providing the foundation for scalable and fault-tolerant distributed applications.

Message-queue (MQ) systems serve as the backbone of this communication infrastructure, decoupling producers from consumers and enabling flexible publish/subscribe interactions across services. Unlike traditional request-response models, message queues absorb traffic bursts, enable temporal decoupling between system components, and facilitate seamless scaling of individual services. Over the past decade, the MQ landscape has diversified substantially: in addition to high-throughput streaming platforms such as Kafka, practitioners increasingly rely on lightweight brokers and protocol stacks like MQTT~\cite{mqtt_spec}, NATS~\cite{nats_io}, Redis Pub/Sub~\cite{redis_pubsub}, and Zenoh~\cite{zenoh_io}, as well as multi-protocol engines such as RabbitMQ~\cite{rabbitmq} and ActiveMQ Artemis~\cite{activemq_artemis}. Each of these systems embodies different trade-offs in architecture, protocol semantics, and runtime implementation---from single-threaded event loops to multi-threaded actor models, from lightweight C implementations to feature-rich Java virtual machines---which can lead to sharply different performance and resource profiles in practice.

Choosing an appropriate message broker for a given deployment is inherently challenging. 
A broker that performs well in a well-provisioned cloud environment may not exhibit the same performance on resource-constrained edge hardware \cite{koziolek2020comparison}. 
Likewise, systems optimized for high throughput can suffer from poor tail latency under bursty workloads \cite{dean2013tail}, while complex broker deployments may impose memory requirements that are challenging for embedded gateways~\cite{luzuriaga2015handling}. These trade-offs are especially pronounced in edge and IoT settings, where deployments span powerful fog nodes to severely constrained microcontrollers, and where network variability, power budgets, and physical space impose strict operational limits~\cite{fahad2025ensemble, lertpongrujikorn2025edgeweaver, paul2026oaas}.

Although prior studies have evaluated message brokers across different environments, they typically focus on a single protocol family (e.g., MQTT~\cite{koziolek2020comparison, mishra2021stress, dizdarevic2024benchmarking}) or a limited subset of systems~\cite{chy2023comparative, liang2023performance}, making it difficult to compare fundamentally different broker designs under comparable conditions. More importantly, existing benchmarks largely emphasize throughput and latency on fixed hardware configurations, overlooking CPU and memory footprint as well as the effects of varying resource allocations. These dimensions are critical for deploying and tuning message brokers in resource-constrained edge and IoT environments. As a result, practitioners lack consistent empirical guidance for selecting and sizing messaging infrastructure for telemetry, streaming analytics, and IoT workloads.

To address this gap, we present mq-bench, a unified benchmarking framework that applies consistent pub/sub workloads to heterogeneous MQ systems and records throughput, latency distributions, and resource consumption in a standardized format. Using mq-bench, we conduct a head-to-head evaluation of eight widely deployed brokers---Mosquitto, EMQX, HiveMQ, RabbitMQ, ActiveMQ Artemis, NATS Server, Redis Pub/Sub, and Zenoh Router---under identical hardware and workload conditions.
Our study focuses on single-node vertical scaling 
and resource efficiency across real-world IoT and edge deployments. We scale experiments from 500 to 10,000 concurrent client pairs and provision brokers with varying resource allocations (1--4 vCPUs, 2--8~GB RAM), spanning the strict constraints of edge devices (e.g., Raspberry Pi) to small cloud instances.
By jointly measuring CPU and memory footprint alongside throughput and latency, our evaluation enables practitioners to accurately size broker deployments and assess resource headroom for target workloads.

In summary, this paper makes the following main contributions:
\begin{itemize}[noitemsep,topsep=0pt]
    \item We introduce \textbf{mq-bench}\footnote{The MQ-Bench source code is available at: \url{https://github.com/hpcclab/mq-bench}}, an extensible, open-source benchmarking tool written in Rust that enables reproducible cross-protocol MQ evaluations with precise latency measurement and resource monitoring.
    \item We present a controlled experimental methodology on dedicated infrastructure, scaling up to 10K concurrent pub-sub pairs and reporting throughput, tail-latency percentiles (p50, p95), and CPU/memory footprint across three VM configurations representative of edge deployments.
    \item We provide a comparative performance characterization of eight brokers, revealing the maximum load each system can sustain under different resource constraints: NATS achieves the highest scalability (90K~msg/s at 10K connections), followed by Zenoh and RabbitMQ, while single-threaded Mosquitto saturates at 45K~msg/s.
    \item We analyze resource efficiency and show that Java-based brokers (HiveMQ, Artemis) consume 10--50$\times$ more memory than native implementations---a critical consideration for memory-constrained edge deployments.
    \item Based on our findings, we provide practical deployment guidelines that map workload requirements and resource constraints to appropriate broker choices, helping practitioners make informed infrastructure decisions.
\end{itemize}

The rest of the paper is organized as follows. Section~\ref{sec:background} provides background on message brokers and the publish/subscribe paradigm. 
Section~\ref{sec:related-work} reviews related work. 
Section~\ref{sec:design} describes our experimental design, including benchmark scenarios, metrics, and testbed infrastructure. 
Section~\ref{sec:results} presents our evaluation results, 
and Section~\ref{sec:conclusion} concludes with key findings and future directions.

%% file: Sources/background.tex
\section{Background}
\label{sec:background}
This section introduces the publish/subscribe messaging model and provides an overview of the messaging protocols and broker architectures evaluated in this study. We focus on single-node deployments to establish a baseline for performance comparison.



\subsection{Publish / Subscribe Architecture}
Message queues enable asynchronous communication between distributed components. In the publish/subscribe model, publishers send messages to named channels called \emph{topics}. Subscribers register interest in specific topics and receive messages published to them. A central \emph{broker} routes messages from publishers to subscribers.

This decoupling offers several advantages. Publishers and subscribers operate independently without knowing each other's identity or location. The broker buffers messages during traffic bursts, preventing slower consumers from blocking faster producers. New publishers or subscribers can join without modifying existing components.

\begin{figure}[htbp]
    \centering
    \includegraphics[width=0.5\textwidth]{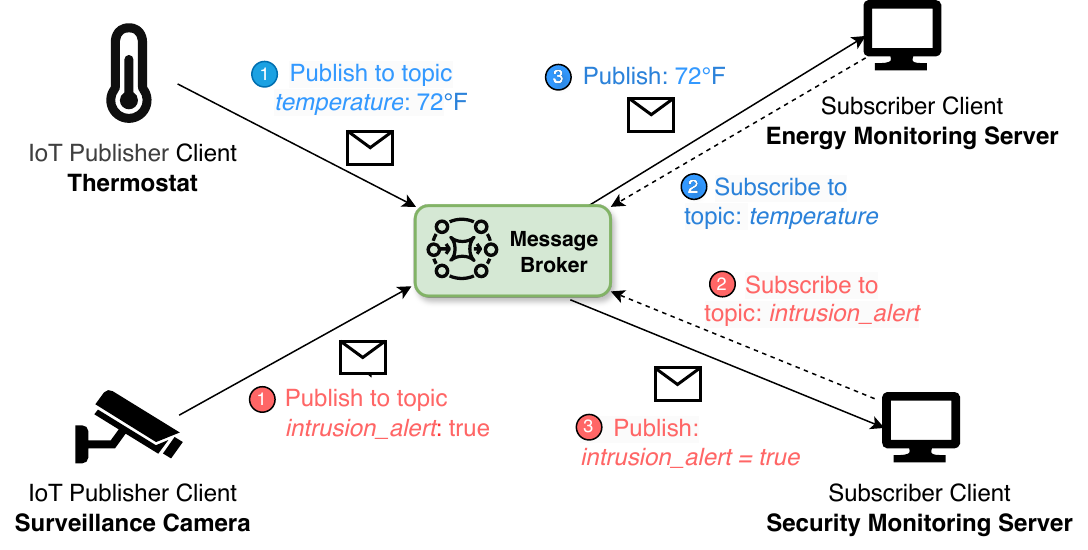}
    \caption{\small Pub/sub architecture for IoT and distributed systems. Publishers send data to named topics; a central broker routes messages to subscribers, decoupling producers from consumers.}
    \label{fig:mq-architecture}
    \vspace{-2mm}
\end{figure}

Figure~\ref{fig:mq-architecture} illustrates a typical pub/sub deployment in IoT systems. Devices such as an electricity meter and thermostat publish sensor data to topics (e.g., \texttt{temperature}, \texttt{intrusion\_alert}), while backend systems subscribe to receive updates. The broker handles all message routing, enabling flexible scaling without direct connections between components.

When evaluating pub/sub systems, several key performance characteristics must be considered:
\begin{itemize}[noitemsep,topsep=2pt]
    \item \textbf{Throughput}: Messages delivered per second under sustained load.
    \item \textbf{Latency}: Time from message publication to subscriber receipt.
    \item \textbf{Scalability}: Ability to handle increasing clients and message rates.
    \item \textbf{Resource efficiency}: CPU and memory consumption relative to workload.
\end{itemize}

These metrics depend on broker architecture, protocol overhead, and runtime environment. Native implementations in C, Rust, or Go typically achieve lower latency and memory usage than managed-runtime brokers in Java or Erlang, though the latter may offer richer features and easier fault tolerance.

\subsection{Messaging Protocols}
\noindent\textbf{ MQTT (Message Queuing Telemetry Transport)}~\cite{mqtt_spec} is a lightweight publish/subscribe protocol for constrained devices and unreliable networks, offering three Quality of Service (QoS) levels (Table~\ref{tab:mqtt_qos}).

\begin{table}[htbp]
\centering
\small
\begin{tabular}{|c|p{1.4cm}|p{5.1cm}|}
\hline
\textbf{QoS} & \textbf{Guarantee} & \textbf{Description} \\
\hline
0 & At most once & Fire-and-forget; no acknowledgment. Fastest but may lose messages. \\
\hline
1 & At least once & Acknowledged delivery; may produce duplicates if ACK is lost. \\
\hline
2 & Exactly once & Four-step handshake ensures no loss or duplication. Highest overhead. \\
\hline
\end{tabular}
\caption{\small MQTT Quality of Service levels. Higher QoS increases reliability at the cost of latency and broker overhead.}
\label{tab:mqtt_qos}
\vspace{-5mm}
\end{table}
\noindent\textbf{AMQP (Advanced Message Queuing Protocol)}~\cite{amqp_spec} is an open standard for business messaging with sophisticated routing (exchanges and queues) and strong reliability guarantees. We use AMQP~0-9-1 for RabbitMQ experiments.

\noindent\textbf{NATS}~\cite{nats_io} is a cloud-native messaging system using a simple text-based protocol. Core NATS offers ``at most once" delivery, prioritizing low latency and high throughput.

\noindent\textbf{Zenoh}~\cite{zenoh_io} is a data-centric protocol blending publish/subscribe with distributed storage. It supports peer-to-peer communication, enabling brokerless or lightweight-router deployments that reduce network overhead.

\noindent\textbf{RESP (Redis Serialization Protocol)}~\cite{redis_pubsub} is the underlying protocol used by Redis. It provides fast pub/sub via an in-memory architecture, achieving low latency but lacking the durability of dedicated message queues.

\subsection{Message Brokers}
We evaluate a diverse set of brokers summarized in Table~\ref{tab:brokers}, categorizing them by their primary architectural focus:

\begin{table}[htbp]
\centering
\small
\begin{tabular}{|l|l|l|}
\hline
\textbf{Broker} & \textbf{Language} & \textbf{Supported Protocols} \\
\hline
Mosquitto & C & MQTT \\
EMQX & Erlang & MQTT \\
HiveMQ CE & Java & MQTT \\
RabbitMQ & Erlang & AMQP~0-9-1, MQTT \\
ActiveMQ Artemis & Java & MQTT \\
NATS Server & Go & NATS \\
Redis & C & RESP \\
Zenoh Router & Rust & Zenoh \\
\hline
\end{tabular}
\caption{\small Message brokers evaluated in this study. Only primary protocols relevant to our study are listed; some brokers support additional protocols (e.g., STOMP, OpenWire).}
\label{tab:brokers}
\vspace{-1.8mm}
\end{table}

\subsubsection{MQTT-Centric Brokers}
These brokers are optimized specifically for the MQTT protocol, ranging from lightweight implementations to massive-scale engines.

\noindent\textbf{Mosquitto}~\cite{mosquitto} is an open-source Eclipse project written in C, using a single-threaded event-loop architecture optimized for resource-constrained hardware.

\noindent\textbf{EMQX}~\cite{emqx} is built on Erlang/OTP, leveraging the actor model for massive concurrency and fault tolerance, targeting carrier-grade MQTT deployments.

\noindent\textbf{HiveMQ}~\cite{hivemq} is an enterprise-grade, Java-based MQTT broker using a non-blocking, asynchronous architecture for high scalability.

\subsubsection{Enterprise Message Brokers}
These systems prioritize reliability, complex routing, and protocol flexibility over raw speed, often serving as the backbone for business applications.

\noindent\textbf{RabbitMQ}~\cite{rabbitmq} is a widely used Erlang-based broker for AMQP 0-9-1, offering flexible routing and strong delivery guarantees, often at the cost of higher latency.

\noindent\textbf{ActiveMQ Artemis}~\cite{activemq_artemis} is a high-performance, non-blocking Java message broker that supports multiple protocols, including AMQP, MQTT, and OpenWire.

\subsubsection{Cloud-Native \& Data-Centric Architectures}
These systems depart from traditional messaging models to address specific needs like microservices communication, or edge computing.

\noindent\textbf{NATS Server}~\cite{nats_io} is a CNCF (Cloud Native Computing Foundation) project written in Go, prioritizing simplicity and performance. Its ``fire-and-forget" architecture minimizes overhead, making it ideal for high-frequency microservices communication.

\noindent\textbf{Zenoh}~\cite{zenoh_io} is written in Rust with a data-centric architecture. It operates as a router or in brokerless peer-to-peer mode, representing edge-native protocols for the cloud-to-edge continuum.

\noindent\textbf{Redis}~\cite{redis_pubsub}, although primarily an in-memory key-value database, is widely used for lightweight messaging. We include its Pub/Sub mechanism as a performance baseline against dedicated brokers.

%% file: Sources/related-work.tex
\section{Related Studies}
\label{sec:related-work}

\begin{table*}[h!]
\centering
\renewcommand{\arraystretch}{1.3}

\resizebox{\textwidth}{!}{%
\begin{tabular}{|l|c|c|c|c|c|c|c|c|c|c|c|c|c|c|c|}
\hline
    References
    & \textbf{Protocol}
    & \multicolumn{8}{c|}{\textbf{Message Queues}}
    & \multicolumn{6}{c|}{\textbf{Metrics}} \\
\hline
&
& \textbf{Mosq.}
& \textbf{EMQX}
& \textbf{HiveMQ}
& \textbf{Artemis}
& \textbf{RabbitMQ}
& \textbf{NATS}
& \textbf{Redis}
& \textbf{Zenoh}
& \textbf{Latency}
& \textbf{Throughput}
& \textbf{Resource}
& \textbf{Scalability}
& \textbf{QoS}
& \textbf{Other} \\
\hline

Liang et al. (2023) \cite{liang2023performance}
& M, Z
& \checkmark &  &  &  &  &  &  & \checkmark
& \checkmark & \checkmark &  &  &  &  \\ \hline

Koziolek et al. (2020) \cite{koziolek2020comparison}
& M
&  & \checkmark & \checkmark &  &  &  &  &
& \checkmark & \checkmark & \checkmark &  &  & \checkmark \\ \hline

Dizdarevic et al. (2023) \cite{dizdarevic2024benchmarking}
& M
& \checkmark & \checkmark & \checkmark & \checkmark &  &  &  &
& \checkmark &  &  &  &  & \checkmark \\ \hline

Mishra (2018) \cite{mishra2018performance}
& M
& \checkmark &  & \checkmark & \checkmark &  &  &  &
& \checkmark & \checkmark &  &  & \checkmark & \checkmark \\ \hline

Mishra et al. (2021) \cite{mishra2021stress}
& M
& \checkmark & \checkmark & \checkmark &  &  &  &  &
& \checkmark & \checkmark & \checkmark &  & \checkmark &  \\ \hline

Pazos et al. (2024) \cite{pazos2024performance}
& M
& \checkmark & \checkmark & \checkmark &  &  &  &  &
& \checkmark &  &  &  & \checkmark & \checkmark \\ \hline

Ibrahim et al. (2025) \cite{ibrahim2025impact}
& A, N
&  &  &  & \checkmark & \checkmark & \checkmark &  &
& \checkmark & \checkmark &  & \checkmark &  &  \\ \hline

Maharjan et al. (2023) \cite{maharjan2023benchmarking}
& A, R
&  &  &  & \checkmark & \checkmark &  & \checkmark &
& \checkmark & \checkmark &  &  &  &  \\ \hline

Chy et al. (2023) \cite{chy2023comparative}
& A
&  &  &  & \checkmark &  &  &  &
& \checkmark & \checkmark & \checkmark &  &  & \checkmark \\ \hline

Our Study
& M, A, N, R, Z
& \checkmark & \checkmark & \checkmark & \checkmark & \checkmark & \checkmark & \checkmark & \checkmark
& \checkmark & \checkmark & \checkmark & \checkmark & \checkmark &  \\ \hline
\end{tabular}
} 
\caption{Comparison of message queue systems and evaluation metrics in existing studies. Protocol: M=MQTT, A=AMQP, N=NATS, R=Redis, Z=Zenoh.}
\label{tab:mq-comparison}
\vspace{-6mm}
\end{table*}

\noindent\textbf{MQTT-Focused Benchmarks.}
The majority of messaging system benchmarks concentrate on MQTT within IoT contexts~\cite{mishra2021stress, dizdarevic2024benchmarking, kashyap2024implementation, pazos2024performance, koziolek2020comparison, mishra2018performance}. These studies primarily evaluate popular MQTT brokers such as Mosquitto, EMQX, and HiveMQ, examining key performance indicators including message throughput, end-to-end latency, edge deployment scenarios, QoS level behavior, and horizontal clustering capabilities. Despite their valuable contributions, these studies remain protocol-locked to MQTT and do not consider alternative messaging paradigms.

\noindent\textbf{Enterprise and Cloud-Native Message Queues.}
A separate line of research investigates enterprise-grade message brokers~\cite{ibrahim2025impact, maharjan2023benchmarking, chy2023comparative}, comparing systems such as RabbitMQ, ActiveMQ Artemis, and NATS using standardized benchmarking frameworks. However, these enterprise-focused studies typically omit MQTT brokers entirely, creating a gap in understanding how IoT-oriented and enterprise systems compare under equivalent conditions.

\noindent\textbf{Emerging Protocols and Data-Centric Middleware.}
Recent work has begun exploring newer messaging paradigms beyond traditional broker architectures. Liang et al.~\cite{liang2023performance} studied Zenoh in Industrial IoT settings, highlighting potential latency benefits relative to MQTT-based systems. In addition, Redis Pub/Sub remains relatively understudied in academic literature despite its widespread adoption in production systems for lightweight messaging.

\noindent\textbf{Gap and Contribution.}
As Table~\ref{tab:mq-comparison} summarizes, no existing study comprehensively evaluates MQTT brokers (Mosquitto, EMQX, HiveMQ), multi-protocol message queues (RabbitMQ, ActiveMQ Artemis), cloud-native systems (NATS), in-memory pub/sub transports (Redis), and data-centric middleware (Zenoh) within a unified experimental framework. Our work addresses this gap through mq-bench, a Rust-based benchmarking tool that enables reproducible, cross-protocol performance comparisons using consistent workload patterns, measurement methodologies, and resource utilization metrics across eight broker systems.

%% file: Sources/experimental-design.tex
\section{Experimental Design}
\label{sec:design}

Message brokers are often deployed in resource-constrained edge environments where they must handle high client loads and varying message sizes. Our evaluation focuses on understanding how different broker architectures perform under these challenging conditions. Specifically, we aim to answer:
\begin{itemize}[noitemsep,topsep=2pt]
    \item \textbf{Latency under varying payloads}: How does message size affect end-to-end delivery latency? IoT applications range from small sensor readings to bulk data transfers, and brokers must handle this diversity efficiently.
    \item \textbf{Throughput under high client load}: What is the maximum number of concurrent clients each broker can sustain? As deployments scale, brokers face increasing connection counts and message rates.
    \item \textbf{Resource efficiency under constraints}: How do brokers perform when CPU and memory are limited? Edge devices often have strict resource budgets, making efficiency critical.
    \item \textbf{Reliability under network instability}: How do MQTT brokers maintain QoS guarantees when network conditions are unstable?
\end{itemize}

To address these objectives, we design three experiments which are detailed in the (Section~\ref{sec:scenarios}). We describe the metrics collected in Section~\ref{sec:metrics}, broker deployment configurations in Section~\ref{sec:broker-config}, workload generation strategy in Section~\ref{sec:workload}, the benchmark tool in Section~\ref{sec:tool}, and testbed infrastructure in Section~\ref{sec:testbed}.

\subsection{Benchmark Scenarios}
\label{sec:scenarios}
\begin{figure}[htbp]
    \centering
    \includegraphics[width=0.45\textwidth]{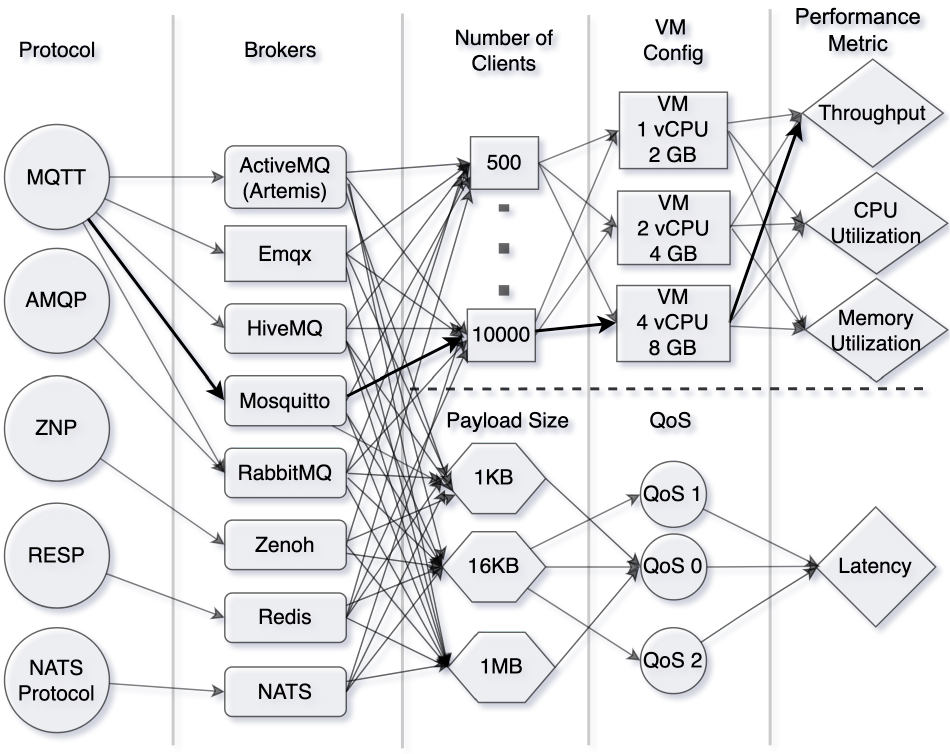}
    \vspace{-2mm}
    \caption{\small Experimental scenarios evaluated in this study. Each path from start to end represents one test scenario. For instance, the solid black path represents the scenario in which MQTT–Mosquitto is evaluated with 10,000 clients on a VM with 4-vCPU, 8-GB RAM configuration while measuring throughput.}
    \label{fig:test-coverage}
    \vspace{-7mm}
\end{figure}

We design three experiments to evaluate broker performance. Figure~\ref{fig:test-coverage} illustrates the test scenarios.

\subsubsection{Experiment 1: Latency vs. Payload Size}
This experiment characterizes how message size affects end-to-end latency. We measure median (p50) and tail (p95) latency at three payload sizes: 1~KB (sensor telemetry), 16~KB (IoT aggregation), and 1~MB (bulk data transfer). Table~\ref{tab:exp1_params} summarizes the parameters.

\vspace{-2mm}

\begin{table}[htbp]
\centering
\small
\begin{tabular}{|l|l|}
\hline
\textbf{Parameter} & \textbf{Value} \\
\hline
Number of pub-sub pairs & 10 \\
Message rate (per publisher) & 10 msg/s \\
Payload sizes & 1 KB, 16 KB, 1 MB \\
Test duration & 120 seconds \\
Warmup period & 60 seconds \\
QoS level & 0 (best-effort) \\
\hline
\end{tabular}
\caption{\small Latency vs. Payload Size parameters.}
\label{tab:exp1_params}
\vspace{-8mm}
\end{table}

\subsubsection{Experiment 2: Throughput and Resource Utilization}
This experiment measures broker capacity and resource consumption under increasing client load. We progressively increase the number of pub-sub pairs until saturation. Table~\ref{tab:exp2_params} summarizes the parameters.

We measure throughput, CPU utilization, and memory consumption at each scale point. Saturation is indicated by throughput dropping below the offered load. We repeat measurements across three VM configurations (Section~\ref{sec:vm-config}) to characterize vertical scaling efficiency.

In addition to the 1-to-1 topology, we repeat this experiment with fanout topology.
\vspace{-2mm}
\begin{table}[htbp]
\centering
\small
\begin{tabular}{|l|l|}
\hline
\textbf{Parameter} & \textbf{Value} \\
\hline
Number of pub-sub pairs & 500, 1000, 2000, ..., 10,000 \\
Message rate (per publisher) & 10 msg/s \\
Offered load & \#pairs $\times$ rate \\
Payload size & 1 KB \\
Test duration & 120 seconds \\
Warmup period & 60 seconds \\
QoS level & 0 (best-effort) \\
\hline
\end{tabular}
\caption{\small Throughput and Resource Utilization parameters.}
\label{tab:exp2_params}
\vspace{-3mm}
\end{table}

\subsubsection{Experiment 3: QoS Reliability Under Network Failures}
This experiment evaluates how MQTT brokers maintain Quality of Service guarantees when subscribers experience network failures. Unlike controlled disconnections, network failures via TCP RST packets simulate realistic cloud and edge scenarios where connections drop unexpectedly. Table~\ref{tab:exp3_params} summarizes the parameters.

With MTTF=30s and MTTR=5s over a 180-second duration, each test experiences approximately 5--6 network failures with a total downtime of 25--30 seconds (approximately 85\% uptime). We test only the five MQTT-native brokers (Mosquitto, EMQX, HiveMQ, RabbitMQ, and Artemis) as NATS, Redis, and Zenoh use different QoS semantics. For each QoS level, we measure message loss percentage and end-to-end latency to characterize both reliability and the latency impact of different QoS levels.

\begin{table}[htbp]
\centering
\small
\begin{tabular}{|l|l|}
\hline
\textbf{Parameter} & \textbf{Value} \\
\hline 
Number of pub-sub pairs & 10 \\
Message rate (per publisher) & 10 msg/s \\
Payload size & 1 KB \\
Test duration & 180 seconds \\
Mean time to failure (MTTF) & 30 seconds \\
Mean time to recovery (MTTR) & 5 seconds \\
\hline
\end{tabular}
\caption{\small{QoS Reliability Under Network Failures parameters.}}
\vspace{-6mm}
\label{tab:exp3_params}
\end{table}

\subsection{Measurement Metrics}
\label{sec:metrics}
We collect the following metrics during each experiment:

\begin{itemize}
    \item \textbf{Latency}: End-to-end message latency measured from send timestamp (embedded in message header) to receive timestamp (recorded by subscriber). We report the minimum, median (p50) and tail-latency percentile (p95) to capture both typical and worst-case behavior.
    
    \item \textbf{Throughput}: The number of messages successfully delivered per second, calculated as $\text{Throughput} = {\text{received\_count}}/{\text{duration}}$ (message per second).
    
    \item \textbf{CPU Utilization}: Broker container CPU usage as a percentage of allocated vCPUs, sampled via Docker \cite{docker} Stats API at 1-second intervals.
    
    \item \textbf{Memory Footprint}: Broker container resident set size (RSS) in megabytes, sampled via Docker Stats API.
\end{itemize}

\subsection{Broker Deployment Configurations}
\label{sec:broker-config}

\subsubsection{Broker Selection}
We evaluate the eight message broker implementations summarized in Table~\ref{tab:brokers} (Section~\ref{sec:background}), covering lightweight MQTT brokers (Mosquitto), enterprise-grade MQTT platforms (EMQX, HiveMQ), multi-protocol message queues (RabbitMQ, ActiveMQ Artemis), cloud-native systems (NATS), in-memory pub/sub (Redis), and data-centric middleware (Zenoh). For Zenoh, we strictly configure it in client/router mode rather than peer-to-peer mode to ensure a fair comparison with other broker-based systems. All brokers are deployed as single-node instances without clustering to isolate core broker performance from distributed coordination overhead. 
To ensure fair comparison, we configure each broker to fully utilize the available CPU resources by tuning thread pool sizes and worker counts if needed to match the number of allocated vCPUs. The exception is Mosquitto, which employs a single-threaded event-loop architecture and cannot be parallelized across multiple cores. 
\subsubsection{Container and VM Configuration}
\label{sec:vm-config}
Each broker runs inside a Docker container \cite{denninnart2023efficiency}, \cite{chanikaphon2023ums} hosted within a KVM-based virtual machine on the Broker Execution Machine. Docker containers are configured with high file descriptor limits (\texttt{nofile: 300000}) to support large numbers of concurrent connections. For scalability experiments, we use three VM configurations that mirror typical edge computing hardware such as the Raspberry Pi~\cite{raspberry_pi_wiki}: 
    (1) \textbf{1~vCPU / 2~GB RAM}: Constrained edge devices such as IoT gateways,
    (2) \textbf{2~vCPU / 4~GB RAM}: Mid-range edge hardware such as small industrial PCs.
    (3) \textbf{4~vCPU / 8~GB RAM}: Higher-capacity fog nodes and small servers.

\subsubsection{Protocol Mappings}
Each broker is tested using its native protocol: MQTT for Mosquitto, EMQX, and HiveMQ; NATS for NATS Server; RESP for Redis; and Zenoh for Zenoh Router. ActiveMQ Artemis is evaluated via its MQTT interface for protocol-level consistency. RabbitMQ is evaluated using both MQTT and AMQP~0-9-1 in separate runs.

\vspace{-2mm}

\subsection{Workload Generation Strategy}
\label{sec:workload}

\subsubsection{Communication Topology}
We adopt a 1-to-1 (pair) topology where each publisher is paired with exactly one subscriber on a dedicated topic. This design provides isolation (no inter-client interference), enables scalability testing (stress connection handling and routing), ensures reproducibility (symmetric setup), and establishes a baseline for future fan-in/fan-out studies. For the throughput experiment, we additionally evaluate a fanout topology on the 4~vCPU, 8~GB RAM VM, where a single publisher broadcasts to $N$ subscribers on a shared topic.

\subsubsection{QoS Configuration}
First two experiments use QoS level 0 (at-most-once delivery) for MQTT-based brokers and equivalent best-effort semantics for other protocols. This ensures fair comparison across protocols and measures peak broker capacity. The third experiment, however, evaluates all three MQTT QoS levels.

\subsection{Testbed Overview}
\label{sec:testbed}

\begin{figure}[htbp]
    \centering
    \includegraphics[width=0.4\textwidth]{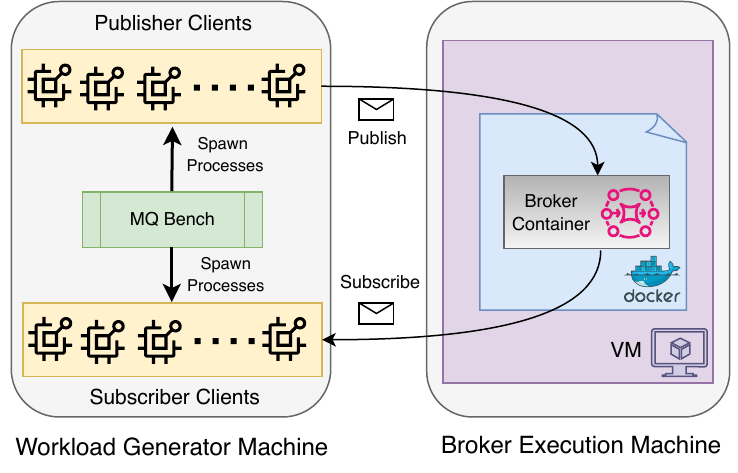}
    \caption{\small{Experimental testbed architecture. The Workload Generator Machine runs benchmarking tool (mq-bench) to spawn publisher/subscriber clients that send and receive messages to/from the broker container hosted on the Broker Execution Machine within a VM.}}
    \label{fig:testbed}
    \vspace{-4mm}
\end{figure}

Figure~\ref{fig:testbed} illustrates the architecture of our experimental testbed, which consists of two dedicated bare-metal nodes provisioned from Chameleon Cloud \cite{chameleoncloud} (node type: \texttt{compute\_icelake\_r650}, image: \texttt{CC-Ubuntu22.04}), connected via a high-speed private network.

\textbf{Workload Generator Machine.} This machine hosts the mq-bench benchmarking tool and orchestrates the entire experiment lifecycle. At the start of each iteration, the tool brings up the broker container on the remote Broker Execution Machine. It then spawns concurrent Publisher and Subscriber clients that generate load against the broker according to the configured workload parameters. During execution, mq-bench collects all transmitted and received messages along with their timestamps, which are used to calculate performance metrics including throughput, latency, and message loss. Upon completion, the tool brings the broker container down before proceeding to the next iteration. A detailed description of mq-bench's architecture is provided in Section~\ref{sec:tool}.

\textbf{Broker Execution Machine.} This machine provides an isolated execution environment for the message broker under 
test. The broker runs within a Docker container~\cite{docker},~\cite{chanikaphon2023ums} inside a KVM-based virtual machine, ensuring consistent resource
allocation and minimal interference from host processes. The hypervisor layer hosts three pre-configured VM instances as mentioned in the Section~\ref{sec:vm-config} to evaluate broker scalability under varying computational resources. 
A resource monitoring module continuously tracks CPU and memory footprint of the broker container in real-time throughout each experiment iteration.

\subsection{Benchmark Tool}
\label{sec:tool}

\begin{figure}[htbp]
    \centering
    \includegraphics[width=0.8\linewidth]{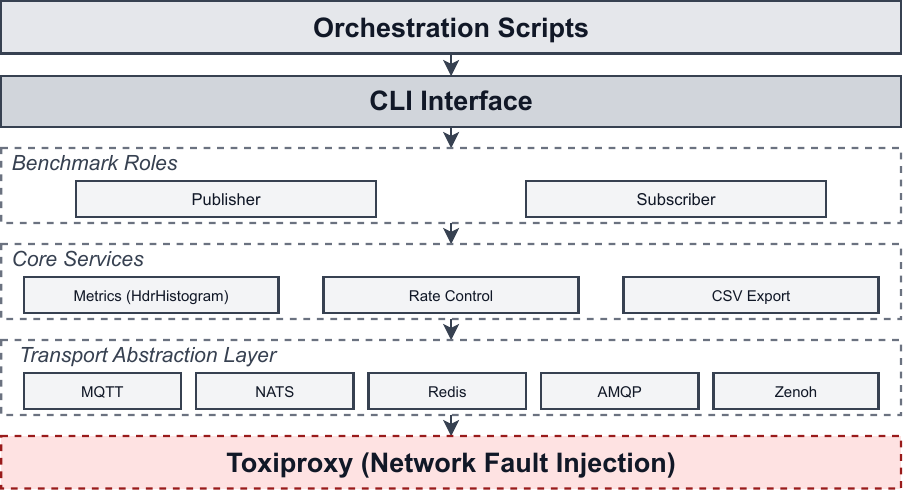}
    \caption{\small Architecture of mq-bench. Orchestration scripts invoke the CLI to spawn Publisher and Subscriber roles. Core services handle metrics collection, rate limiting, and result export.}
    \label{fig:mq-bench-arch}
    \vspace{-4mm}
\end{figure}

As introduced in Section~\ref{sec:testbed}, we developed \textbf{mq-bench}, a high-performance benchmarking tool that resides on the workload generator machine (Figure~\ref{fig:testbed}) and orchestrates the entire experiment lifecycle. Developed in Rust using the Tokio asynchronous runtime, mq-bench utilizes a transport abstraction layer to offer a unified load-generation interface across various messaging protocols. Figure~\ref{fig:mq-bench-arch} depicts the tool's architecture.

\subsubsection{Architecture}
The benchmark tool consists of pluggable transport adapters for MQTT (via \texttt{rumqttc}), NATS (via \texttt{async-nats}), Redis (via \texttt{redis-rs}), AMQP (via \texttt{lapin}), and Zenoh (native SDK). Each adapter implements a common \texttt{Transport} trait supporting subscribe, publish, and connection management operations.

\subsubsection{Latency Measurement}
Each message includes a 24-byte binary header containing a 64-bit send timestamp in UNIX nanoseconds. Upon receiving a message, subscribers extract this timestamp and compute end-to-end latency as $\text{latency} = t_{\text{recv}} - t_{\text{send}}$, where $t_{\text{recv}}$ is the local receive time. Clock synchronization between publisher and subscriber processes is ensured by running both on the same workload generator machine. We record latency values in nanoseconds and aggregate samples to report percentiles, average, and standard deviation.

\subsubsection{Throughput Measurement}
Publishers operate in open-loop mode using a token-bucket rate controller, emitting messages at a fixed rate (e.g., 10~msg/s per publisher) independent of broker acknowledgments or backpressure. This design ensures a consistently offered load regardless of broker response times, accurately revealing saturation behavior when brokers cannot keep pace. To accurately measure throughput, we identify a stable period during each experiment run. A post-processing script analyzes the connection logs and detects when all expected clients have successfully connected to the broker. In case of broker overload and the broker cannot accept target client connections, the script instead identifies when the connection count reaches saturation. Throughput is then calculated only from messages received during this stable period, excluding the initial connection phase. This approach ensures that measurements reflect steady-state broker performance rather than transient startup behavior.

\subsubsection{Network Fault Injection}
For resilience testing, mq-bench integrates with Toxiproxy~\cite{toxiproxy}, a TCP proxy that intercepts connections and injects configurable faults such as latency, bandwidth limits, or connection resets. Failure timing follows an exponential distribution, modeling Poisson failure arrivals common in reliability engineering. The MTTF parameter controls the average interval between failures, while MTTR defines the reconnection delay. Upon proxy recovery, mq-bench automatically reconnects subscribers using the same client ID, enabling MQTT session resumption and message redelivery for QoS 1/2.


%% file: Sources/evaluation.tex
\section{Experimental Results}
\label{sec:results}

This section presents the experimental results from our benchmark study. We evaluate the brokers across three dimensions: latency under varying payload sizes, throughput under client scaling, and reliability under network disruptions. For latency experiments, we use a 4~vCPU, 8GB RAM VM to ensure sufficient headroom and isolate protocol-level differences. For throughput experiments, we vary VM configurations (1, 2, and 4~vCPUs) to observe how brokers scale with available resources.
 
\subsection{Latency vs Payload Size}

\begin{figure}[htbp]
    \centering
    \includegraphics[width=0.43\textwidth]{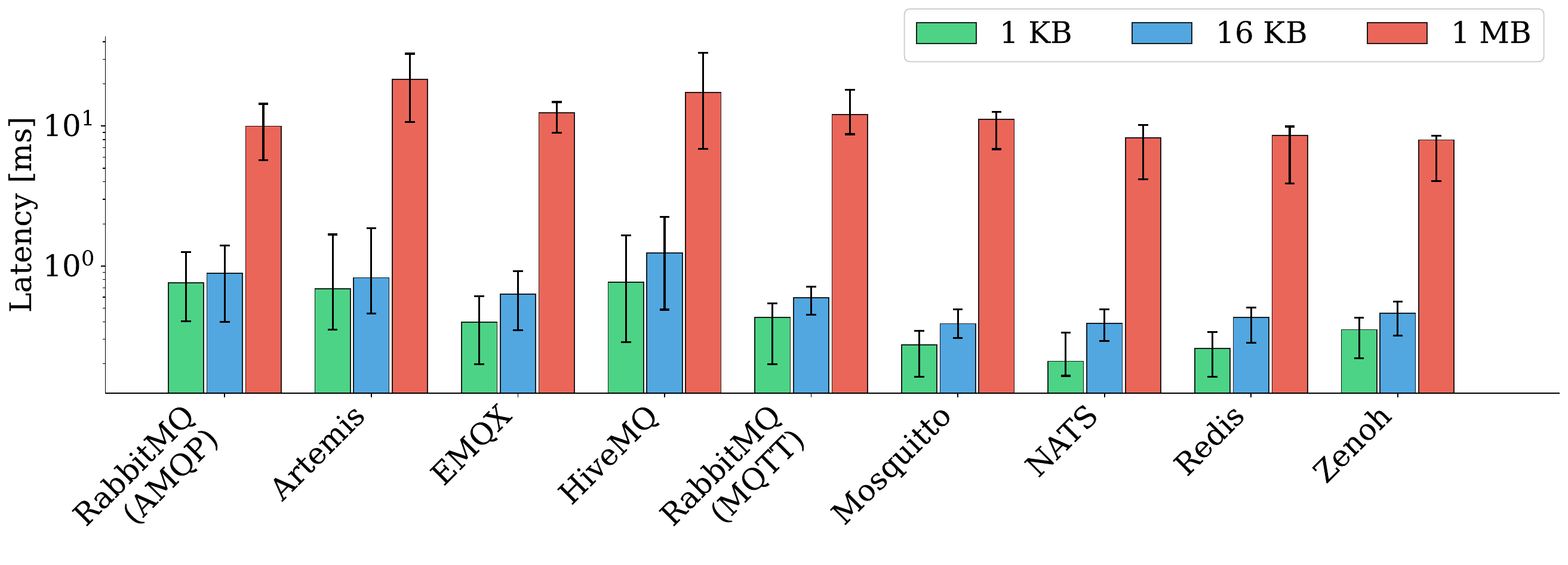}
    \vspace{-3mm}
    \caption{\small{Median latency with error bars (min to P95) across payload sizes. The y-axis uses a logarithmic scale.}}
    \vspace{-4mm}
    \label{fig:latency-vs-payload}
\end{figure}

Figure~\ref{fig:latency-vs-payload} presents end-to-end latency measurements across three payload sizes: 1KB, 16KB, and 1MB. The bar height represents median (P50) latency, with error bars extending from the minimum observed latency to P95.

For small 1KB payloads, native brokers achieve sub-millisecond latencies. NATS leads with the lowest median at 0.21ms, followed closely by Redis (0.26ms) and Mosquitto (0.27ms). Zenoh also performs well at 0.35ms. These brokers benefit from lightweight protocols and efficient memory handling in their native implementations. The managed-runtime brokers show higher latencies: EMQX and RabbitMQ-MQTT both achieve around 0.40--0.43ms, while Artemis and HiveMQ exhibit notably higher medians of 0.69ms and 0.76ms respectively. The JVM-based brokers (HiveMQ, Artemis) show wider variance, with HiveMQ's P95 reaching 1.65ms---roughly 5$\times$ higher than NATS.

At 16KB, all brokers show modest latency increases as expected from the larger serialization and transmission overhead. NATS maintains the lowest median at 0.39ms, with Mosquitto close behind at 0.39ms. The relative ordering remains consistent, though the gap between native and managed-runtime brokers widens slightly. HiveMQ's median increases to 1.24ms with P95 at 2.25ms, reflecting the JVM's additional overhead when processing larger payloads.

For 1MB payloads, serialization and network transfer costs dominate the latency profile. NATS achieves the lowest median at 8.27ms, followed by Zenoh (7.96ms) and Redis (8.58ms). Interestingly, AMQP-RabbitMQ outperforms MQTT-RabbitMQ at this payload size (9.99ms vs 12.11ms), suggesting AMQP's binary framing handles large messages more efficiently. Mosquitto shows higher latency (11.18ms) than expected for a native broker, likely due to its single-threaded architecture becoming a bottleneck during large payload processing. The JVM-based brokers struggle significantly: HiveMQ and Artemis exhibit medians of 17.41ms and 21.60ms respectively, with P95 latencies exceeding 32ms. This 2--3$\times$ latency penalty reflects garbage collection overhead and object serialization costs in managed runtimes when handling large messages. EMQX performs better among managed brokers at 12.42ms, benefiting from Erlang's efficient binary handling.

\begin{tcolorbox}[colback=blue!10, colframe=blue!10, breakable, boxrule=0pt, arc=0pt, left=2pt, right=2pt, top=2pt, bottom=2pt]
\underline{\textbf{Takeaway}:} \emph{For small messages, all brokers deliver similar latencies. As payload size grows, native brokers stay fast while managed-runtime brokers slow down significantly.}
\end{tcolorbox}

\subsection{Throughput vs Client Scaling}

\begin{figure*}[!t]
    \centering
    \includegraphics[width=1.0\textwidth]{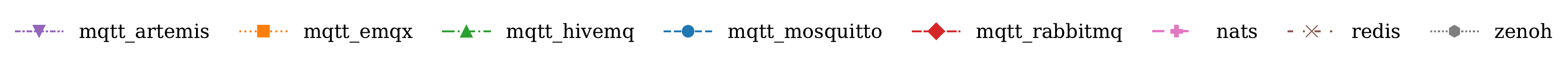}
    
    
    \begin{subfigure}[b]{0.32\textwidth}
        \centering
        \includegraphics[width=\textwidth]{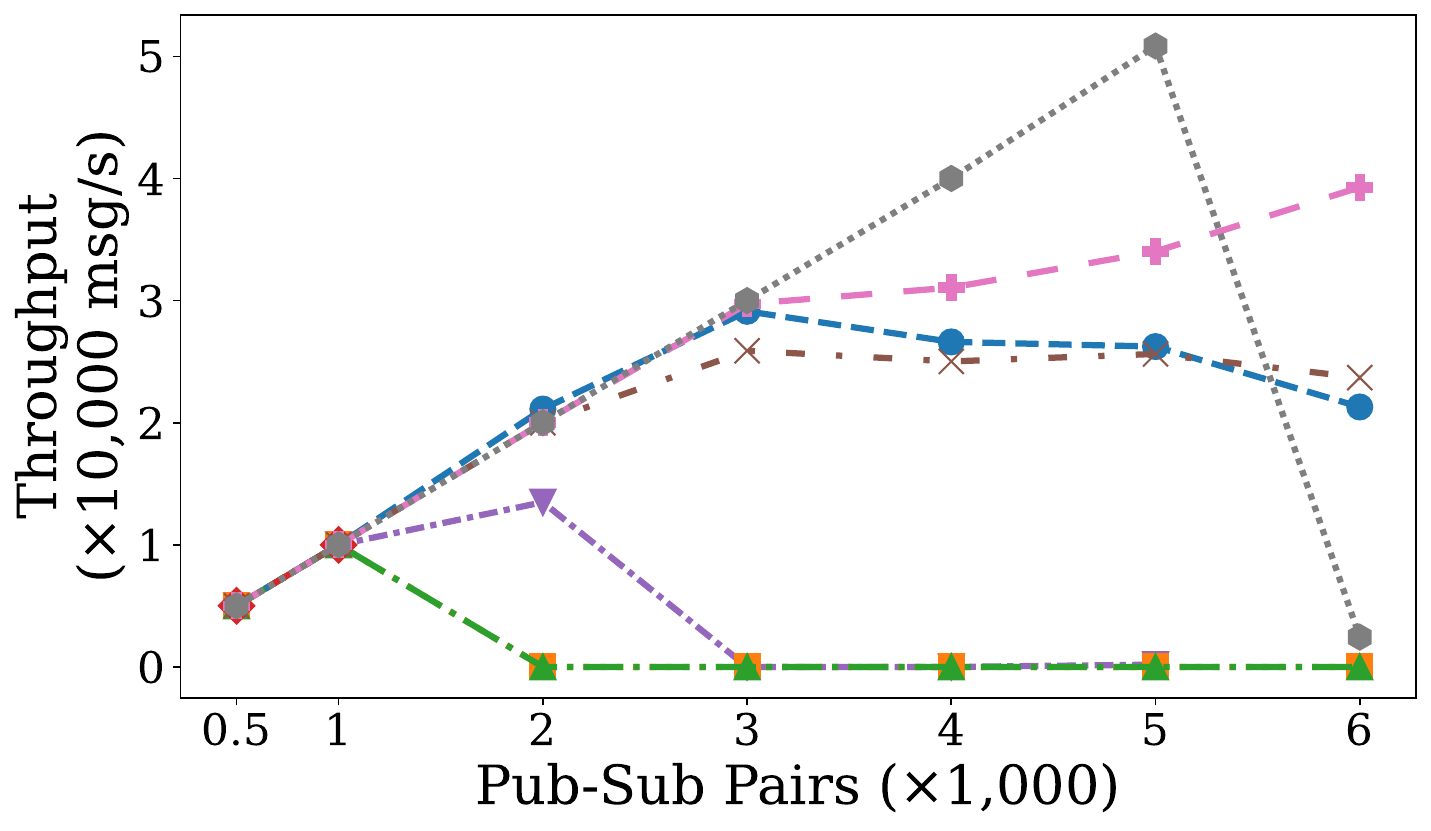}
        \caption{Throughput (1 vCPU, 2GB)}
        \label{fig:throughput-vm1}
    \end{subfigure}
    \hfill
    \begin{subfigure}[b]{0.32\textwidth}
        \centering
        \includegraphics[width=\textwidth]{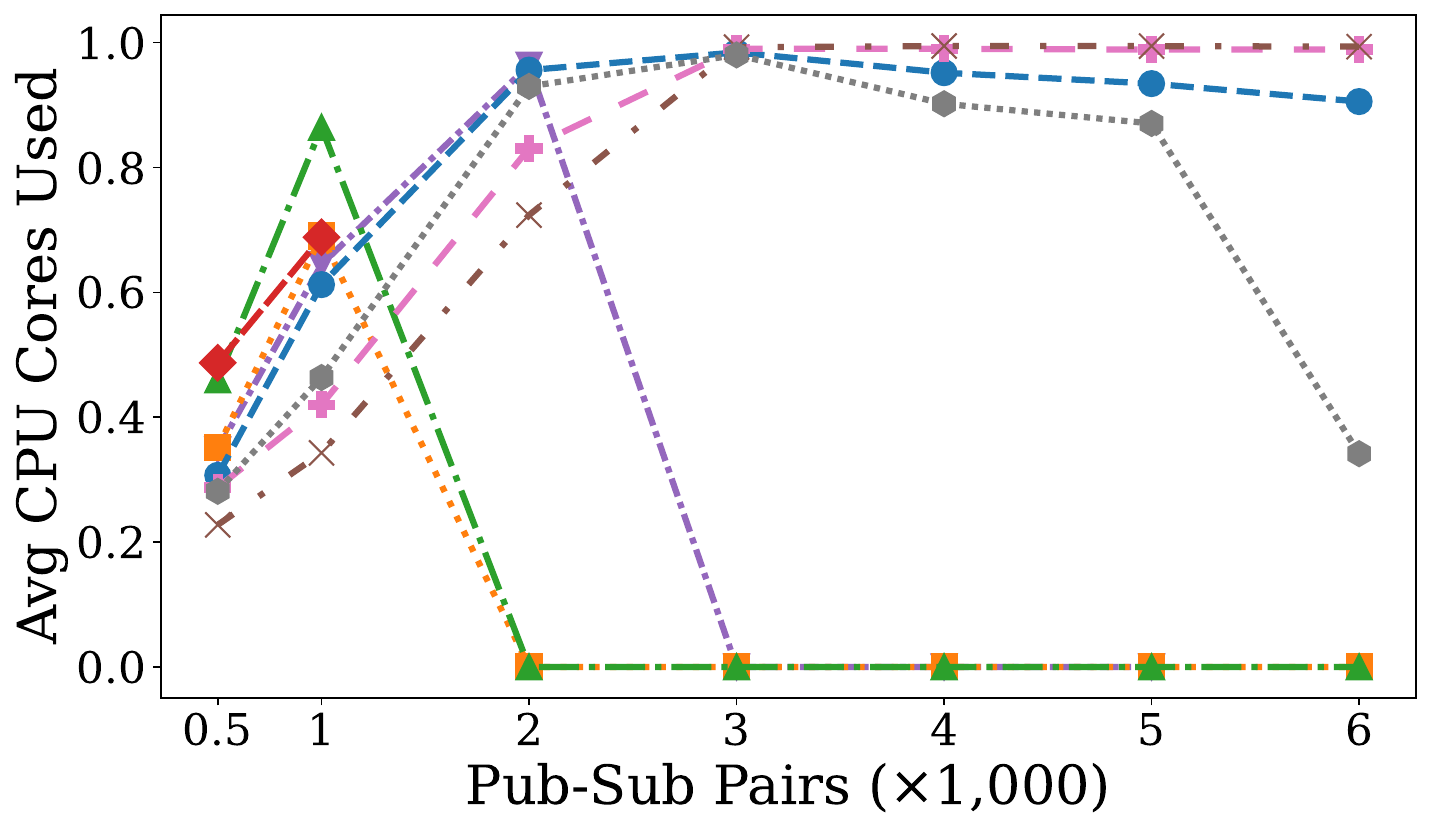}
        \caption{CPU Utilization (1 vCPU, 2GB)}
        \label{fig:cpu-vm1}
    \end{subfigure}
    \hfill
    \begin{subfigure}[b]{0.32\textwidth}
        \centering
        \includegraphics[width=\textwidth]{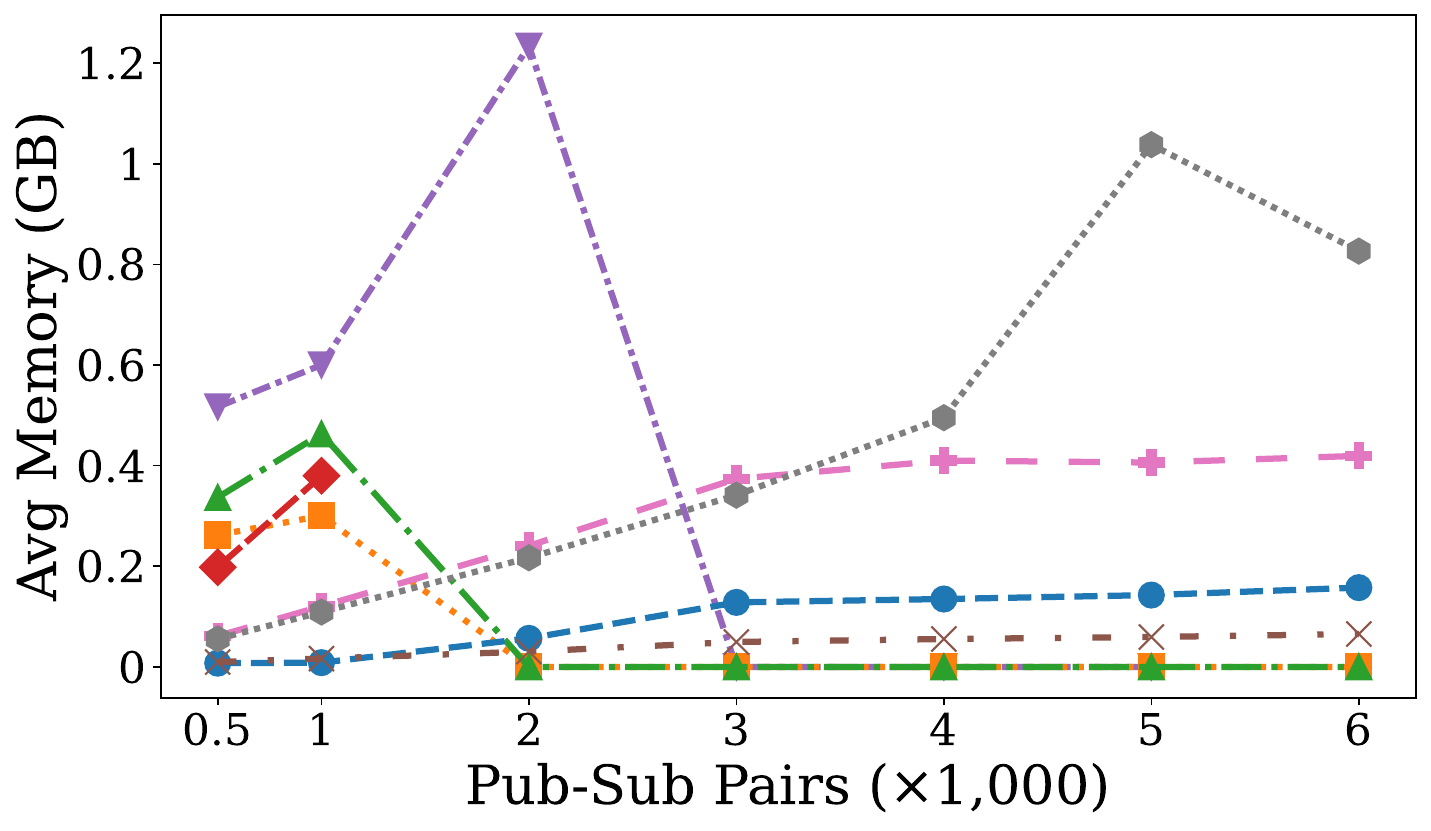}
        \caption{Memory Footprint (1 vCPU, 2GB)}
        \label{fig:memory-vm1}
    \end{subfigure}
    
    \vspace{0.3em}
    
    \begin{subfigure}[b]{0.32\textwidth}
        \centering
        \includegraphics[width=\textwidth]{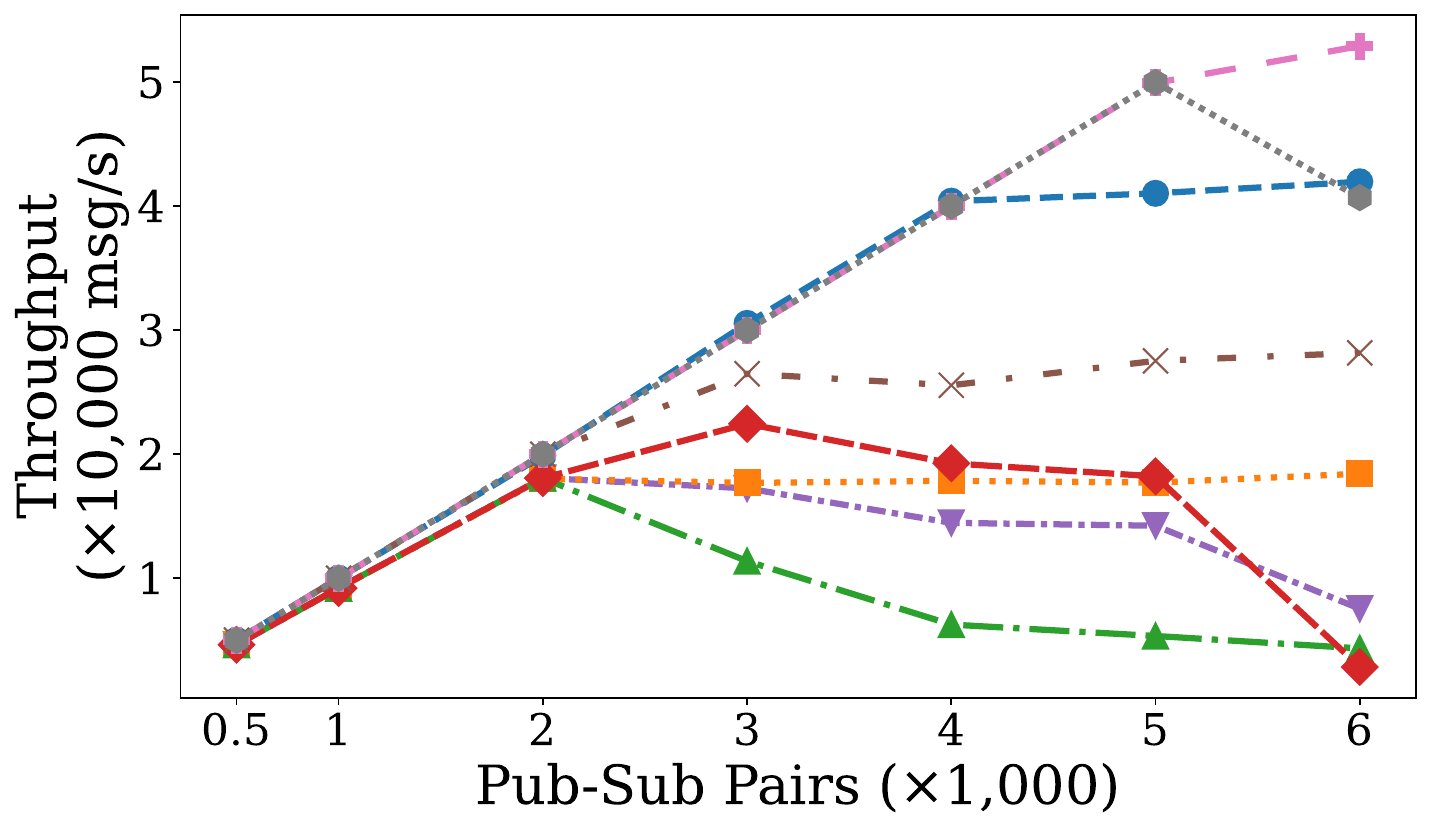}
        \caption{Throughput (2 vCPU, 4GB)}
        \label{fig:throughput-vm2}
    \end{subfigure}
    \hfill
    \begin{subfigure}[b]{0.32\textwidth}
        \centering
        \includegraphics[width=\textwidth]{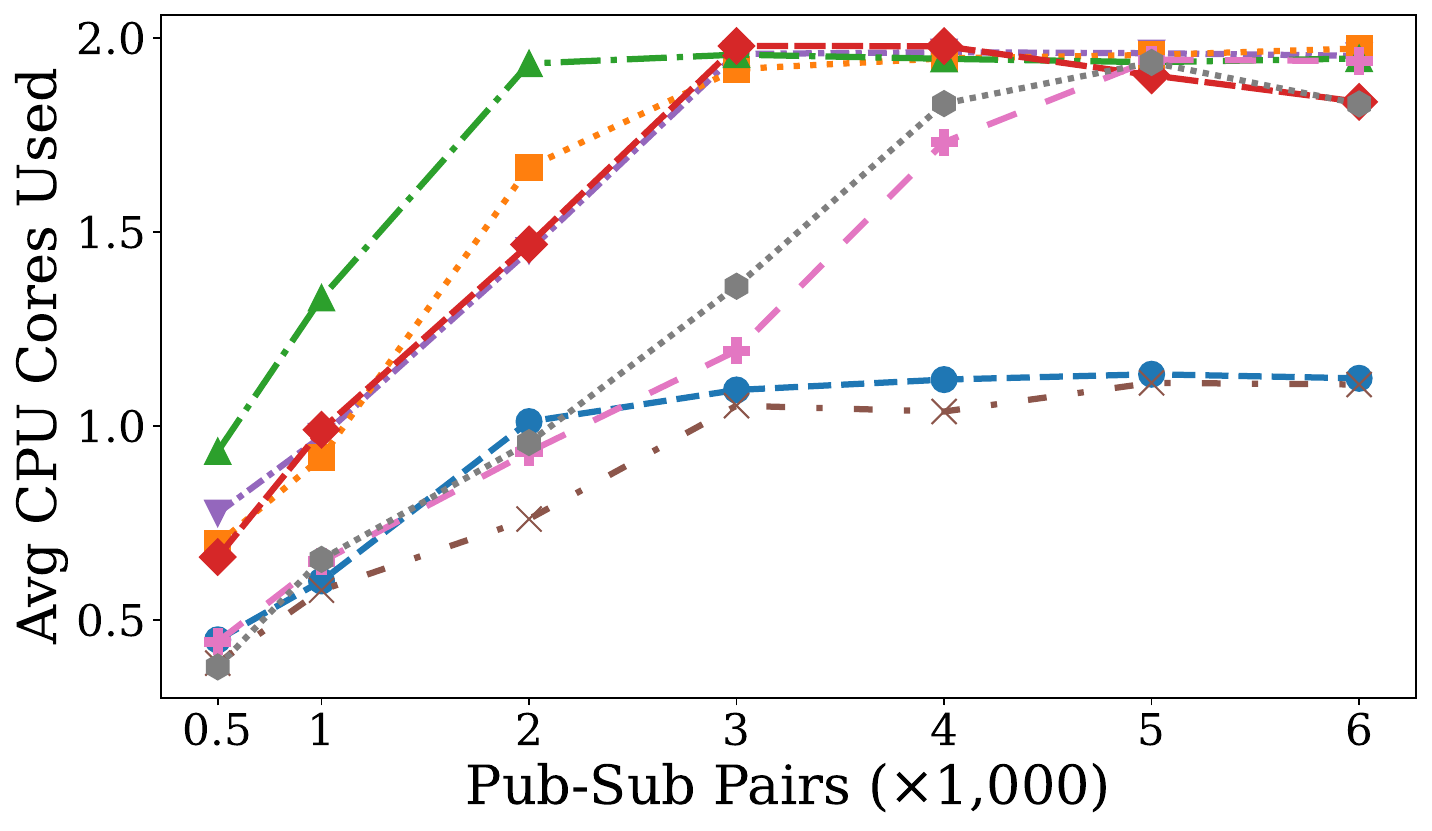}
        \caption{CPU Utilization (2 vCPU, 4GB)}
        \label{fig:cpu-vm2}
    \end{subfigure}
    \hfill
    \begin{subfigure}[b]{0.32\textwidth}
        \centering
        \includegraphics[width=\textwidth]{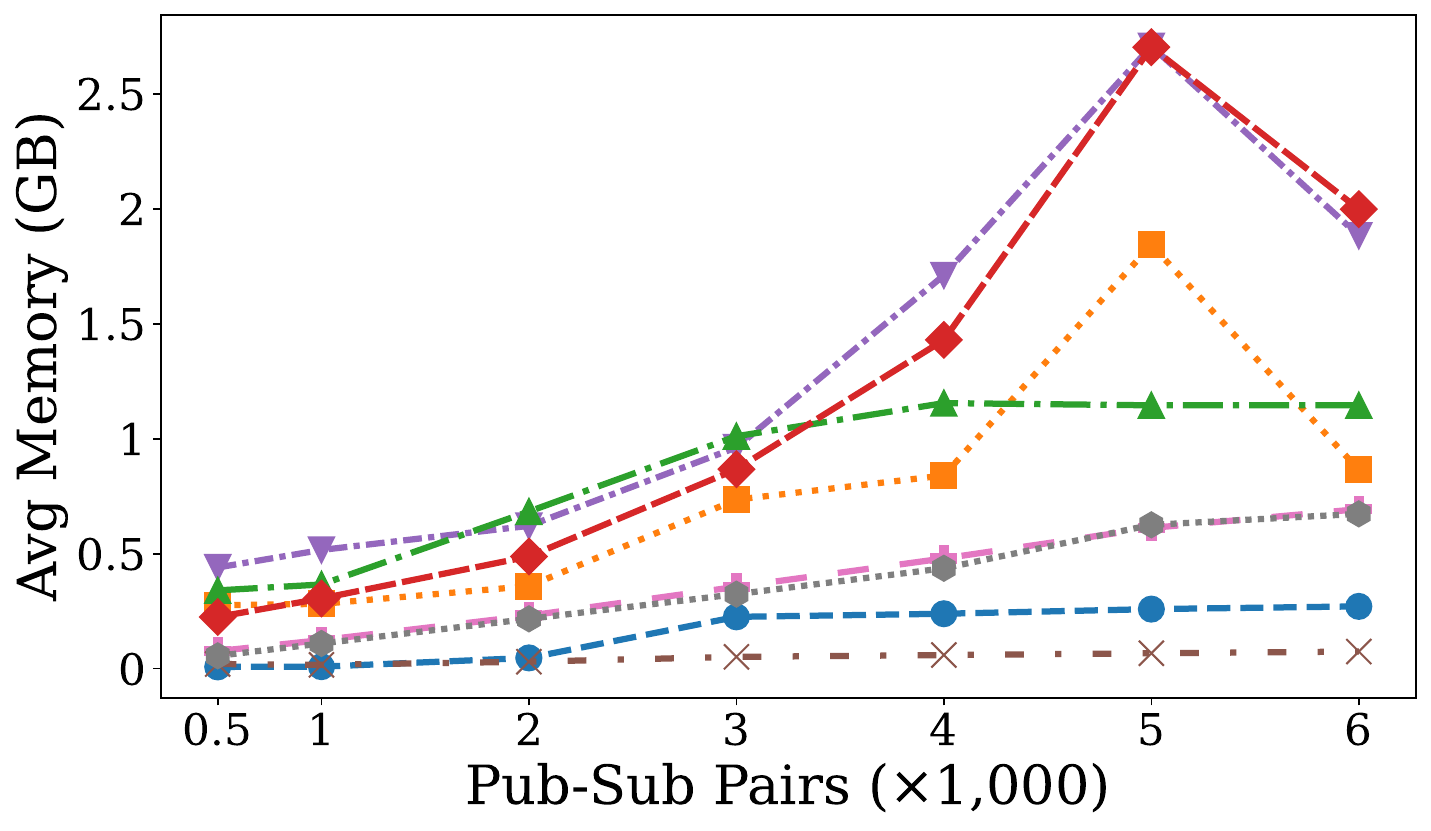}
        \caption{Memory Footprint (2 vCPU, 4GB)}
        \label{fig:memory-vm2}
    \end{subfigure}
    
    \vspace{0.3em}
    
    \begin{subfigure}[b]{0.32\textwidth}
        \centering
        \includegraphics[width=\textwidth]{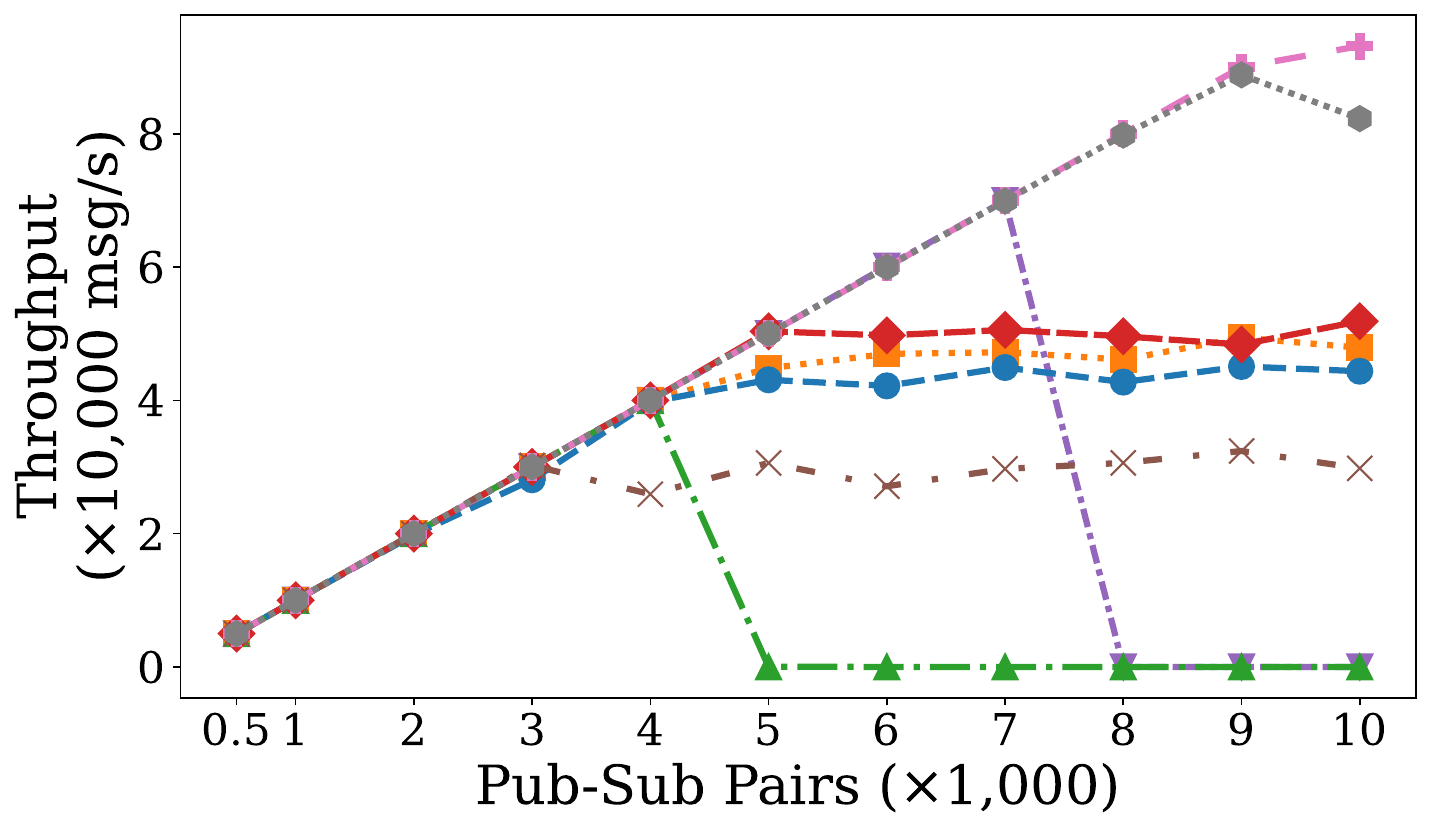}
        \caption{Throughput (4 vCPU, 8GB)}
        \label{fig:throughput-vm4}
    \end{subfigure}
    \hfill
    \begin{subfigure}[b]{0.32\textwidth}
        \centering
        \includegraphics[width=\textwidth]{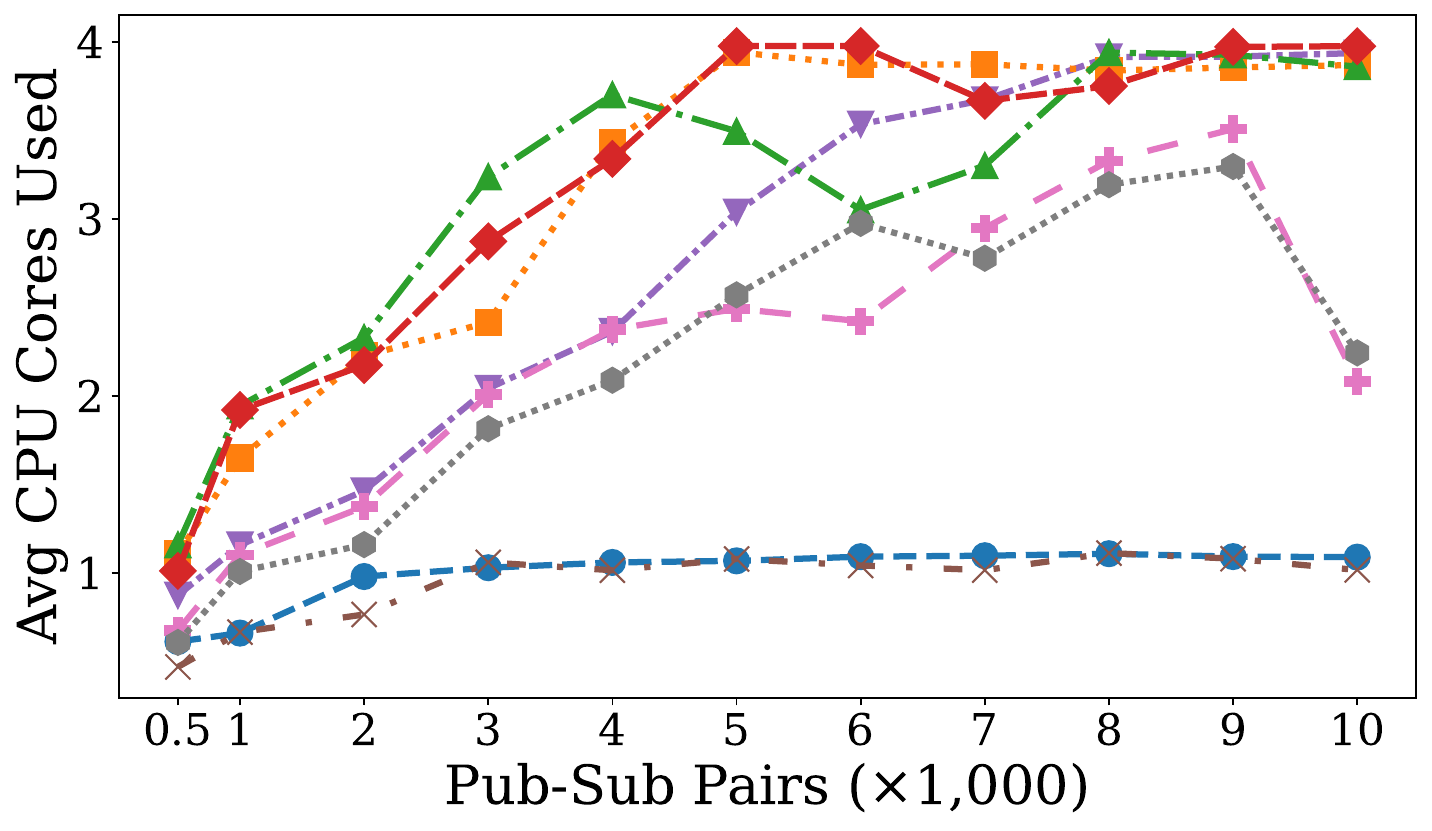}
        \caption{CPU Utilization (4 vCPU, 8GB)}
        \label{fig:cpu-vm4}
    \end{subfigure}
    \hfill
    \begin{subfigure}[b]{0.32\textwidth}
        \centering
        \includegraphics[width=\textwidth]{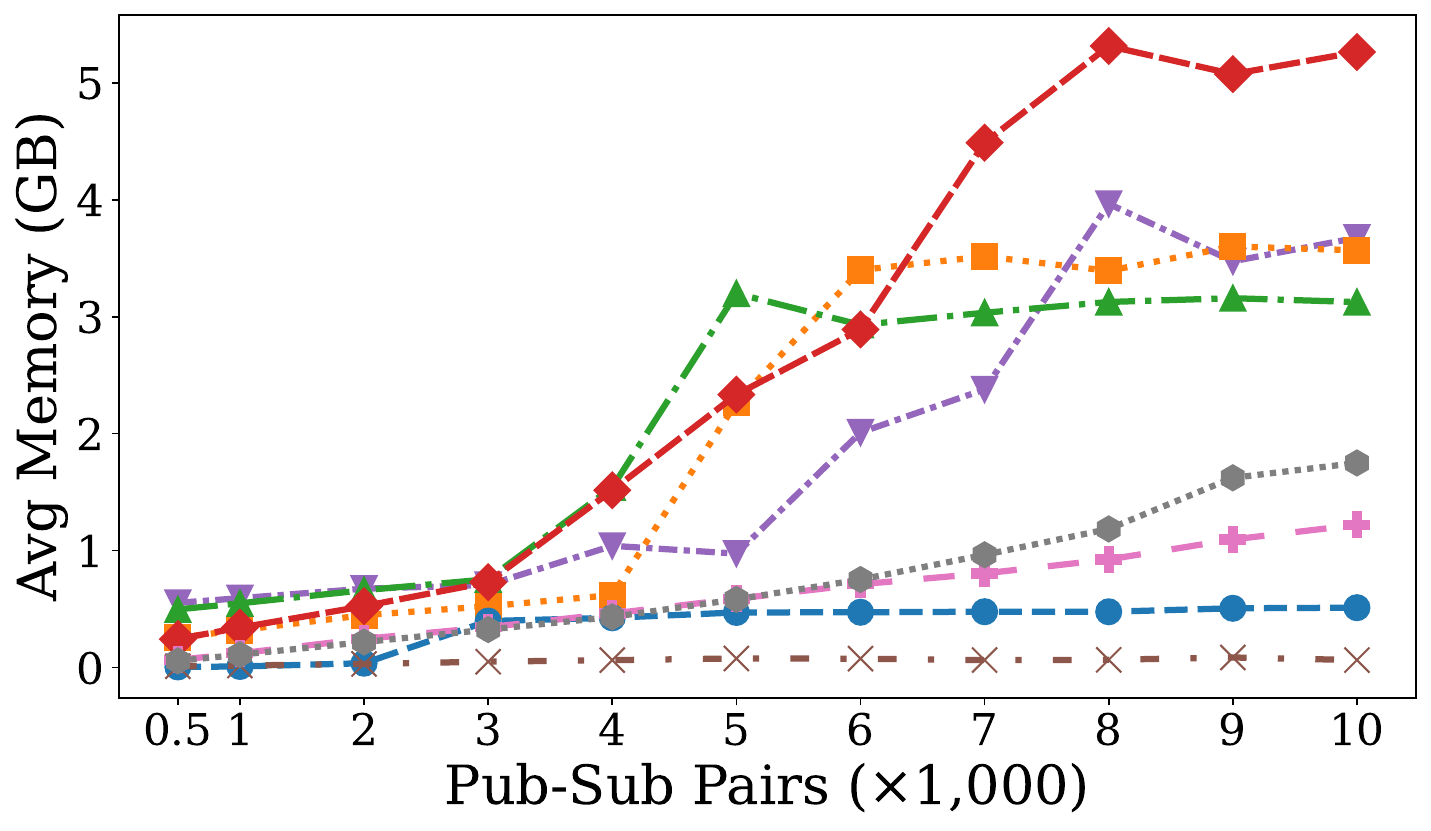}
        \caption{Memory Footprint (4 vCPU, 8GB)}
        \label{fig:memory-vm4}
    \end{subfigure}
    
    \caption{\small Throughput and resource utilization as client pairs scale across three VM configurations. Each row represents a VM configuration (top: 1 vCPU/2GB, middle: 2 vCPU/4GB, bottom: 4 vCPU/8GB). Columns show throughput (msg/s), CPU utilization (cores used), and memory footprint (MB) respectively.}
    \label{fig:throughput-all}
    \vspace{-6mm}
\end{figure*}

This section presents throughput, CPU, and memory footprint results across three VM configurations. With each publisher sending 10~msg/s, target throughput is $\text{pairs} \times 10$ (e.g., 10K~msg/s for 1000 pairs). Figure~\ref{fig:throughput-all} presents all results in a consolidated view. Note that RabbitMQ-AMQP is excluded from the throughput graphs as it failed early in each configuration---after 500 pairs for all the VM configurations. This early failure is attributed to AMQP's heavier protocol overhead compared to MQTT, which requires more resources for connection management and channel multiplexing under high concurrency. Consequently, we were unable to collect throughput data for RabbitMQ-AMQP at higher client pair counts.

\subsubsection{1 vCPU, 2GB RAM}

On our most constrained setup, up to 1000 pairs all brokers keep up without issue. Performance differences emerge as we push higher. Zenoh and NATS pull ahead of the pack---Zenoh sustains an impressive 50.8K msg/s at 5000 pairs, while NATS tops out around 34K msg/s. Both degrade gracefully rather than crashing when pushed beyond their limits. Redis holds its own up to about 25K msg/s before hitting CPU saturation, keeping memory usage minimal at just 62--75MB even under heavy load. Mosquitto's single-threaded architecture caps it at around 10K msg/s, but it remains stable within that range.

The JVM-based brokers---HiveMQ and Artemis---struggle on this limited hardware. HiveMQ stalls beyond 10K msg/s, and Artemis fails entirely past 30K msg/s. Both consume significant memory (487MB and 1.3GB respectively), reflecting the overhead of garbage collection under pressure, and are susceptible to OOM kills in this constrained environment. EMQX hits a wall at 20K msg/s, suggesting the Erlang VM needs more headroom to operate reliably. In some cases, brokers crash shortly after clients begin connecting. When this occurs, all metrics (throughput, CPU, and memory) drop to near-zero in the graphs since we report averages over the entire experiment duration. For instance, HiveMQ \& EMQX crashes at 2000 pairs and Artemis at 3000 pairs. This indicates that brokers written in systems languages like C, Rust, and Go deliver better throughput per resource on constrained hardware than those running on managed runtimes.

\subsubsection{2 vCPU, 4GB RAM}

The additional resources provide substantial improvements for multi-threaded brokers, while single-threaded implementations show more modest gains. Zenoh and NATS benefit significantly from the second core. Zenoh sustains approximately 50K msg/s at 5000 pairs and gracefully degrades to 41K msg/s at 6000 pairs, compared to crashing in VM1. NATS scales to 53K msg/s at 6000 pairs---a 56\% improvement over VM1's 34K msg/s. Both effectively utilize both cores, with moderate memory footprints of 656MB and 728MB respectively.

Redis and Mosquitto, constrained by their single-threaded architectures, show incremental gains. Redis improves from 26K to 28K msg/s and Mosquitto from 26K to 40K msg/s as additional memory helps reduce backpressure. Redis remains memory-efficient at 77--87MB, while Mosquitto uses 284--308MB.

The managed-runtime brokers benefit more noticeably from the doubled memory. EMQX, which failed beyond 1000 pairs in VM1, now sustains 18K msg/s up to 2000 pairs before stalling, consuming 900MB--1.3GB. HiveMQ recovers from complete failure in VM1 to achieve 18K msg/s at 2000 pairs. However, it degrades severely to 4.3K msg/s at 6000 pairs with memory usage of 1.2--1.7GB. RabbitMQ-MQTT reaches 22K msg/s at 3000 pairs before declining to 2.8K msg/s at 6000 pairs, with memory reaching 2.1GB. Artemis degrades from 18K to 7.5K msg/s as load increases, with memory approaching 2GB. Overall, the doubled resources reduce OOM pressure for garbage-collected runtimes, while native multi-threaded implementations gain from parallel execution.

\subsubsection{4 vCPU, 8GB RAM}

With four cores available, multi-threaded brokers scale significantly compared to the VM2. We can scale till 10000 client pairs for most brokers, although the performance varied significantly.
Zenoh can sustain till the 9000 pairs having approximately the perfect throughput of 90K msg/s at 9000 pairs before degrading to 82K msg/s at 10000 pairs.
NATS also reaches 90K msg/s at 9000 pairs. Both utilize 3.3--3.5 cores on average with memory footprints of 1.1--1.7GB.

Redis and Mosquitto, limited by their single-threaded architectures, show minimal gains compared to 2 vCPUs. Redis remains around 30K msg/s, while Mosquitto improves modestly to 45K msg/s. Redis maintains its memory efficiency at 66--90MB, while Mosquitto uses 500--620MB.

The managed-runtime brokers benefit most dramatically from the quadrupled resources. EMQX improves from 18K msg/s in VM2 to 47K msg/s---a 161\% gain---utilizing 3.4--3.9 cores and 650MB--3.7GB memory. Artemis shows the largest improvement, jumping from 18K to 70K msg/s at 7000 pairs before failing at 8000 pairs. RabbitMQ-MQTT sustains 50K msg/s at 5000--6000 pairs, more than doubling its VM2 performance of 22K msg/s. HiveMQ recovers from severe degradation in VM2 to achieve 40K msg/s at 4000 pairs, though it still fails beyond 5000 pairs. These brokers consume substantial memory under load, ranging from 1.6GB to 5.5GB. The results confirm that JVM and Erlang-based brokers require significant resources to perform competitively, but can match or exceed native implementations when adequately provisioned.

\begin{tcolorbox}[colback=blue!10, colframe=blue!10, breakable, boxrule=0pt, arc=0pt, left=2pt, right=2pt, top=2pt, bottom=2pt]
\underline{\textbf{Takeaway}:} \emph{How well a broker scales depends on its threading model and runtime. Multi-threaded native brokers benefit most from extra cores, single-threaded ones hit a ceiling, and managed-runtime brokers need generous resources to keep up.}
\end{tcolorbox}

\subsubsection{Fanout Topology}

\begin{figure}[htbp]
    \centering
    \includegraphics[width=0.4\textwidth]{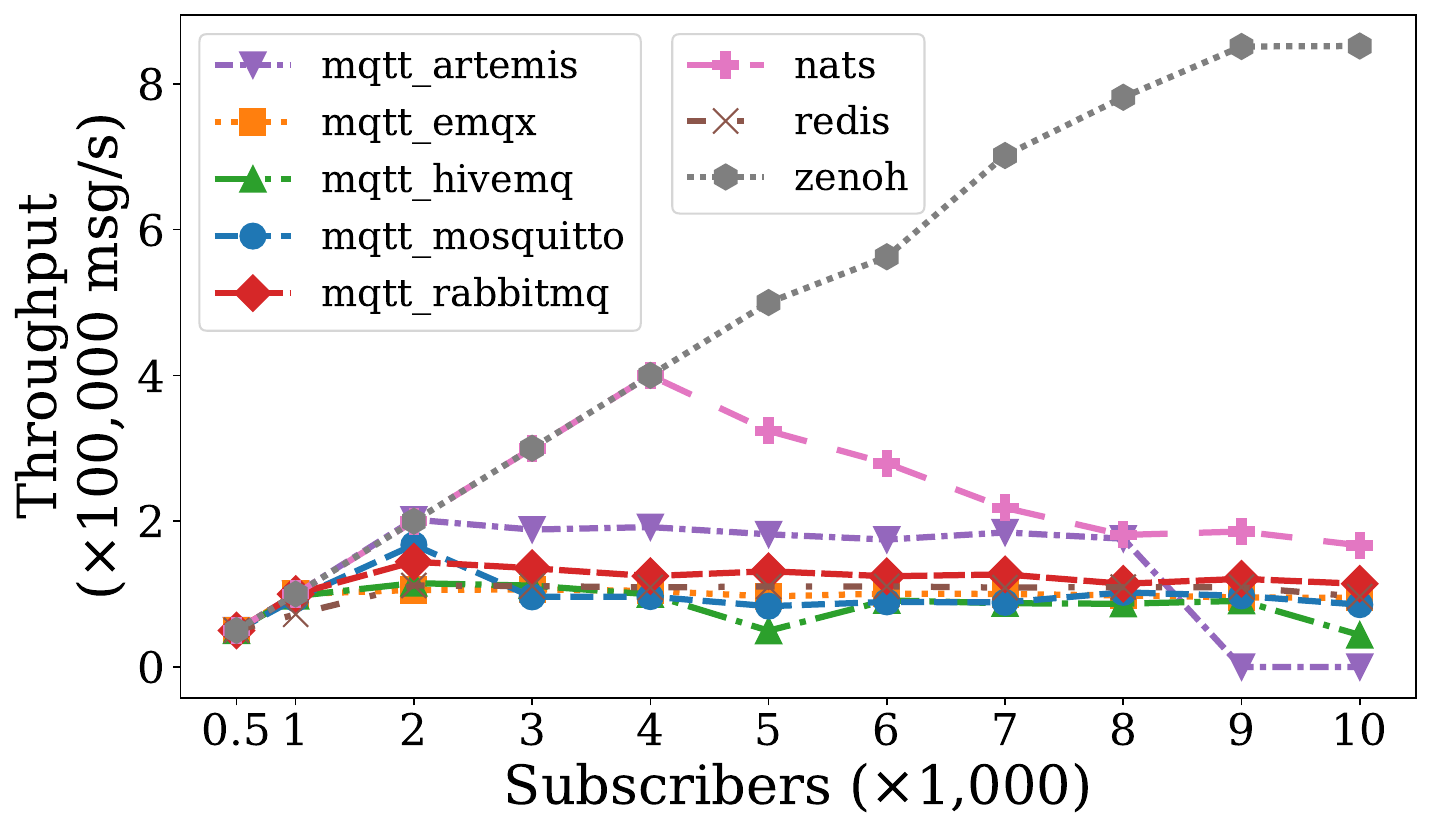}
    \caption{\small{Aggregate subscriber throughput under the fanout topology (1 publisher, $N$ subscribers) on the 4~vCPU, 8GB RAM configuration.}}
    \label{fig:throughput-fanout}
    \vspace{-6mm}
\end{figure}

We additionally evaluate a fanout topology on the 4~vCPU, 8GB RAM VM: a single publisher sends 100~msg/s to a shared topic while $N$ subscribers each receive every message, yielding an aggregate target of $N \times 100$~msg/s. This isolates the broker's dispatch path from its ingestion cost. Note that we use 100~msg/s here instead of the 10~msg/s rate used in the 1-to-1 topology. Because there is only one publisher in the fanout setup, keeping 10~msg/s would place negligible load on the brokers--nearly all of them performed perfectly under that rate--so we increased the message rate to 100~msg/s to stress the dispatch path and expose saturation behavior. Results are shown in Figure~\ref{fig:throughput-fanout}.

Because of this rate difference, raw fanout numbers should not be directly compared with 1-to-1 results. The most striking result is Zenoh's dominance: it sustains up to 850K~msg/s at 10,000 subscribers while CPU plateaus at $\sim$2 cores--well below the 4-core ceiling--suggesting an efficient broadcast dispatch path that avoids per-subscriber overhead. This is a meaningful reversal from 1-to-1, where Zenoh and NATS performed comparably. In fanout, NATS scales well to 4,000 subscribers ($\sim$400K~msg/s) but then collapses to 167K~msg/s at 10,000 despite saturating all four cores, indicating that its dispatch cost scales linearly with subscriber count. Most other brokers--EMQX, Redis, Mosquitto, and HiveMQ--saturate early at $\sim$100K~msg/s regardless of subscriber count, roughly matching their 1-to-1 ceiling. RabbitMQ shows limited scaling in both topologies, while Artemis fails at higher client counts under fanout. CPU and memory usage figures are available in the mq-bench GitHub repository.

\begin{tcolorbox}[colback=blue!10, colframe=blue!10, breakable, boxrule=0pt, arc=0pt, left=2pt, right=2pt, top=2pt, bottom=2pt]
\underline{\textbf{Takeaway}:} \emph{Performance trends from point-to-point topologies generally carry over to fanout, but topology can shift bottlenecks and flip relative rankings: Zenoh, which matched NATS in 1-to-1, dominates in fanout with near-linear throughput scaling and modest CPU usage, while NATS collapses beyond 4,000 subscribers despite saturating all four cores.}
\vspace{-1mm}
\end{tcolorbox}

\vspace{-1mm}
\subsection{Reliability and Latency Under Network Disruptions}
\label{sec:qos-results}

While the previous experiments measured performance under stable conditions, real-world deployments must handle network instability. We evaluate QoS guarantees under TCP RST fault injection, simulating abrupt subscriber disconnections. 

\vspace{1mm}
\noindent
\textbf{Message Reliability Under Network Failures.} Each test transmitted 18,200 messages over 180 seconds while experiencing 5--6 network failure events. At QoS 0 (at-most-once), all brokers exhibit approximately 6.3--6.6\% message loss, corresponding to 1,150--1,200 lost messages. The minor variation across brokers stems from the randomness in failure timing rather than implementation differences, as QoS 0 provides no buffering or redelivery mechanism.

At QoS 1 (at-least-once) and QoS 2 (exactly-once), all five brokers achieve 0\% message loss. This demonstrates that persistent session support (\texttt{clean\_session=false}) combined with stable client identifiers enables reliable message delivery despite multiple network failures. During disconnection periods, brokers buffer messages for offline subscribers and deliver them upon reconnection.

\begin{figure}[htbp]
    \centering
    \includegraphics[width=0.4\textwidth]{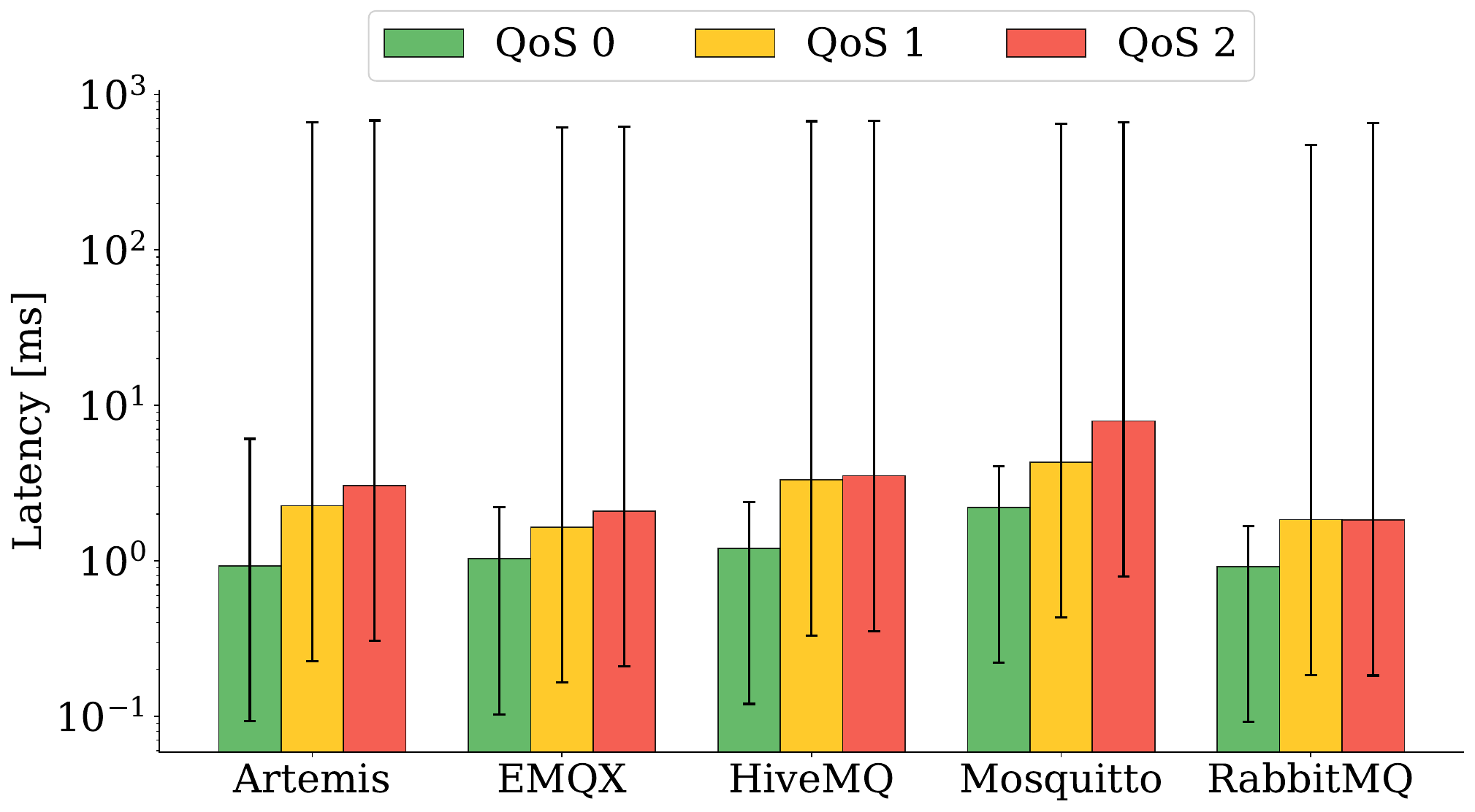}
    \caption{\small Median latency with error bars (min to P95) by broker and QoS level under network failures.}
    \label{fig:qos-latency}
    \vspace{-4mm}
\end{figure}

\vspace{1mm}
\noindent
\textbf{End-to-End Latency Under Failures.} Figure~\ref{fig:qos-latency} presents P50 latency with error bars across QoS levels. The key observation is the dramatic increase in latency variance for QoS 1 and 2 compared to QoS 0.

For QoS 0, all brokers maintain low and consistent latencies with P50 under 2.5ms and P99 under 30ms. Messages are delivered immediately without acknowledgment overhead, resulting in minimal variance.

For QoS 1 and 2, broker architectural differences become apparent. EMQX and RabbitMQ achieve the lowest P50 latencies (1.65--2.1ms), comparable to their QoS 0 performance. Both benefit from lightweight concurrency models: EMQX's Erlang processes and RabbitMQ's actor-based message passing handle acknowledgments asynchronously without blocking the main message path. Artemis and HiveMQ show moderate latencies (2.3--3.5ms), as their JVM thread pools manage acknowledgment state efficiently despite higher per-thread overhead. Mosquitto exhibits the highest latencies, increasing from 2.20ms at QoS 0 to 7.93ms at QoS 2---a 3.6$\times$ penalty. This stems from its single-threaded architecture, which must serialize message delivery with acknowledgment processing, blocking new messages until QoS handshakes complete.

Tail latencies tell a different story: P95 and P99 spike to 475--665ms and $\sim$3,000ms respectively across all brokers uniformly. This reflects message buffering during the 5-second MTTR window, with reconnection overhead dominating rather than broker-specific factors.

\begin{tcolorbox}[colback=blue!10, colframe=blue!10, breakable, boxrule=0pt, arc=0pt, left=2pt, right=2pt, top=2pt, bottom=2pt]
\underline{\textbf{Takeaway}:} \emph{
Higher QoS levels add acknowledgment overhead that multi-threaded brokers handle in the background with little latency cost, while single-threaded brokers must process acknowledgments one at a time, slowing delivery down.
}
\end{tcolorbox}

%% file: Sources/conclusion.tex
\section{Conclusion and Future Works}\label{sec:conclusion}

This paper presented mq-bench, a unified benchmarking framework for evaluating message queue systems across heterogeneous protocols and broker architectures. We evaluated eight brokers--Mosquitto, EMQX, HiveMQ, RabbitMQ, ActiveMQ Artemis, NATS Server, Redis, and Zenoh--under identical hardware and workload conditions.

Our evaluation reveals several key findings. Native brokers written in systems languages (C, Rust, Go) achieve lower latency than managed-runtime brokers, with JVM-based brokers showing 2--3$\times$ higher latency for large payloads. Multi-threaded native brokers (Zenoh, NATS) scale well with additional CPU cores, reaching up to 90K~msg/s, while single-threaded brokers (Redis, Mosquitto) remain stable but do not benefit from parallelism. Memory efficiency varies significantly--Redis maintains minimal usage (66--90MB) while JVM-based brokers consume 10--50$\times$ more, an important consideration for edge deployments.

Our study also uncovered some counterintuitive findings. Managed-runtime brokers can match or exceed native performance when given adequate resources--Artemis achieved 70K~msg/s on 4~vCPU, higher than Redis and Mosquitto. Conversely, Mosquitto showed higher latency than expected for large payloads due to its single-threaded architecture.

\textbf{Future Work.} We plan to extend mq-bench to evaluate distributed multi-node deployments, additional workload patterns such as request-reply semantics, and diverse network conditions typical of wide-area IoT deployments.

Based on our findings, we offer the following guidelines for practitioners:
\begin{tcolorbox}[colback=blue!10, colframe=blue!10, breakable, boxrule=0pt, arc=0pt, left=2pt, right=2pt, top=2pt, bottom=2pt]
\begin{itemize}[noitemsep,topsep=2pt,leftmargin=*]
    \item \textbf{Constrained edge devices:} Single-threaded native brokers (Mosquitto, Redis) offer stability and minimal memory footprint, making them a good fit for resource-constrained deployments.
    \item \textbf{Mid-range edge hardware:} Multi-threaded native brokers (Zenoh, NATS) provide better throughput scaling, well-suited for moderately provisioned edge nodes.
    \item \textbf{Latency-sensitive applications:} Native brokers achieve lower latency, especially for larger payloads, and are preferable for time-critical workloads.
    \item \textbf{Feature-rich deployments:} Managed-runtime brokers (RabbitMQ, Artemis) offer richer features and can deliver competitive performance with sufficient resources.
    \item \textbf{Reliable delivery:} MQTT brokers with QoS 1/2 support reliable delivery; multi-threaded implementations offer lower latency overhead.
\end{itemize}
\end{tcolorbox}

%% file: Sources/acknowledgement.tex
\section*{Acknowledgment}
We would like to thank anonymous reviewers for their constructive feedback; and Chameleon Cloud for providing resources. This project is supported by National Science Foundation (NSF) through CNS CAREER Award\# 2419588.

%% file: references.bib
@article{chy2023comparative,
  title={Comparative evaluation of Java virtual machine-based message queue services: A study on Kafka, Artemis, Pulsar, and RocketMQ},
  author={Chy, Md Showkat Hossain and Arju, Muhammad Ashfakur Rahman and Tella, Sri Manjusha and Cerny, Tomas},
  journal={Electronics},
  volume={12},
  number={23},
  pages={4792},
  year={2023},
  publisher={MDPI}
}

@inproceedings{maharjan2023benchmarking,
  title={Benchmarking message queues},
  author={Maharjan, Rokin and Chy, Md Showkat Hossain and Arju, Muhammad Ashfakur and Cerny, Tomas},
  booktitle={Telecom},
  volume={4},
  
  number={2},
  pages={298--312},
  year={2023},
  organization={MDPI}
}

@article{liang2023performance,
  title={A performance study on the throughput and latency of zenoh, mqtt, kafka, and dds},
  author={Liang, Wen-Yew and Yuan, Yuyuan and Lin, Hsiang-Jui},
  journal={arXiv preprint arXiv:2303.09419},
  year={2023}
}

@inproceedings{ibrahim2025impact,
  title={On the Impact of Message Brokers Implementations in the Choreography of Microservices},
  author={Ibrahim, Ahmed Gamal and Lopes, Rui Pedro and Rufino, Jos{\'e} and Leit{\~a}o, Paulo},
  booktitle={International Conference on Optimization, Learning Algorithms and Applications},
  pages={3--17},
  year={2025},
  organization={Springer}
}

@inproceedings{koziolek2020comparison,
  title={A comparison of MQTT brokers for distributed IoT edge computing},
  author={Koziolek, Heiko and Gr{\"u}ner, Sten and R{\"u}ckert, Julius},
  booktitle={European Conference on Software Architecture},
  pages={352--368},
  year={2020},
  organization={Springer}
}

@article{pazos2024performance,
  title={Performance evaluation of MQTT broker servers deployed in the cloud},
  author={Pazos, Ferando},
  journal={Electronic Journal of SADIO},
  volume={23},
  year={2024}
}

@article{kashyap2024implementation,
  title={Implementation and analysis of EMQX broker for MQTT protocol in the Internet of Things},
  author={Kashyap, Monika and Dev, Ansh Kumar and Sharma, Vidushi},
  journal={e-Prime-Advances in Electrical Engineering, Electronics and Energy},
  volume={10},
  pages={100846},
  year={2024},
  publisher={Elsevier}
}

@misc{mishra2021stress,
  title={Stress-Testing MQTT Brokers: A Comparative Analysis of Performance Measurements. Energies, 14, Article 5817},
  author={Mishra, B and Mishra, B and Kertesz, A},
  year={2021}
}

@inproceedings{dizdarevic2024benchmarking,
  title={Benchmarking Performance of Various MQTT Broker Implementations in a Compute Continuum},
  author={Dizdarevi{\'c}, Jasenka and Michalke, Marc and Jukan, Admela and Masip-Bruin, Xavi and D’Andria, Francesco},
  booktitle={2024 IEEE 24th International Symposium on Cluster, Cloud and Internet Computing (CCGrid)},
  pages={357--366},
  year={2024},
  organization={IEEE}
}

@inproceedings{mishra2018performance,
  title={Performance evaluation of MQTT broker servers},
  author={Mishra, Biswajeeban},
  booktitle={International Conference on Computational Science and Its Applications},
  pages={599--609},
  year={2018},
  organization={Springer}
}

@misc{mqtt_spec,
    author = "OASIS",
    title = "{MQTT Version 5.0}",
    howpublished = {\url{https://docs.oasis-open.org/mqtt/mqtt/v5.0/mqtt-v5.0.html}},
    note="Online; Accessed on 1 Dec. 2025"
}

@misc{amqp_spec,
    author = "OASIS",
    title = "{Advanced Message Queuing Protocol (AMQP) Version 1.0}",
    howpublished = {\url{http://docs.oasis-open.org/amqp/core/v1.0/amqp-core-complete-v1.0.pdf}},
    note="Online; Accessed on 1 Dec. 2025"
}

@misc{nats_io,
    author = "Synadia",
    title = "{NATS Documentation}",
    howpublished = {\url{https://docs.nats.io/}},
    note="Online; Accessed on 1 Dec. 2025"
}

@misc{zenoh_io,
    author = "ZettaScale Technology",
    title = "{Zenoh: Zero Overhead Network Protocol}",
    howpublished = {\url{https://zenoh.io/}},
    note="Online; Accessed on 1 Dec. 2025"
}

@misc{redis_pubsub,
    author = "Redis",
    title = "{Redis Pub/Sub}",
    howpublished = {\url{https://redis.io/docs/interact/pubsub/}},
    note="Online; Accessed on 1 Dec. 2025"
}

@misc{mosquitto,
    author = "Eclipse Foundation",
    title = "{Eclipse Mosquitto}",
    howpublished = {\url{https://mosquitto.org/}},
    note="Online; Accessed on 1 Dec. 2025"
}

@misc{emqx,
    author = "EMQ Technologies",
    title = "{EMQX: The Unified MQTT Platform for AI \& IoT Data Streaming}",
    howpublished = {\url{https://www.emqx.io/}},
    note="Online; Accessed on 1 Dec. 2025"
}

@misc{hivemq,
    author = "HiveMQ",
    title = "{HiveMQ: Enterprise MQTT Broker}",
    howpublished = {\url{https://www.hivemq.com/}},
    note="Online; Accessed on 1 Dec. 2025"
}

@misc{rabbitmq,
    author = "VMware",
    title = "{RabbitMQ: One broker to queue them all}",
    howpublished = {\url{https://www.rabbitmq.com/}},
    note="Online; Accessed on 1 Dec. 2025"
}

@misc{activemq_artemis,
    author = "Apache Software Foundation",
    title = "{ActiveMQ Artemis}",
    howpublished = {\url{https://activemq.apache.org/components/artemis/}},
    note="Online; Accessed on 1 Dec. 2025"
}

@misc{docker,
  title={Docker: Accelerated Container Application Development},
  author={{Docker Inc.}},
  howpublished={\url{https://www.docker.com/}},
  year={2024},
    note="Online; Accessed on 1 Dec. 2025"
}

@misc{iot_analytics_2024,
  title={State of IoT 2024: Number of connected IoT devices growing 13\% to 18.8 billion globally},
  author={{IoT Analytics}},
  howpublished={\url{https://iot-analytics.com/number-connected-iot-devices/}},
  year={2024},
    note="Online; Accessed on 1 Dec. 2025"
}

@article{wu2019wearable,
  title={Wearable technology applications in healthcare: a literature review},
  author={Wu, Min},
  journal={On-Line Journal of Nursing Informatics},
  volume={23},
  number={3},
  year={2019},
  publisher={Healthcare Information and Management Systems Society (HIMSS)}
}

@article{dean2013tail,
  title={The tail at scale},
  author={Dean, Jeffrey and Barroso, Luiz Andr{\'e}},
  journal={Communications of the ACM},
  volume={56},
  number={2},
  pages={74--80},
  year={2013},
  publisher={ACM New York, NY, USA}
}

@inproceedings{luzuriaga2015handling,
  title={Handling mobility in IoT applications using the MQTT protocol},
  author={Luzuriaga, Jorge E and Cano, Juan Carlos and Calafate, Carlos and Manzoni, Pietro and Perez, Miguel and Boronat, Pablo},
  booktitle={2015 Internet Technologies and Applications (ITA)},
  pages={245--250},
  year={2015},
  organization={IEEE}
}

@inproceedings{fahad2025ensemble,
  title={Ensemble learning based wifi sensing using spatially distributed tx-rx links},
  author={Fahad, Nafeez and Touhiduzzaman, Md and Bulut, Eyuphan},
  booktitle={2025 International Conference on Computing, Networking and Communications (ICNC)},
  pages={606--611},
  year={2025},
  organization={IEEE}
}

@misc{toxiproxy,
  title={Toxiproxy: A TCP proxy to simulate network and system conditions for chaos and resiliency testing},
  author={{Shopify}},
  year={2024},
  howpublished={\url{https://github.com/Shopify/toxiproxy}},
  note="Online; Accessed on 1 Dec. 2025"
}

@incollection{chameleoncloud,
  title={Lessons Learned from the Chameleon Testbed},
  author={Kate Keahey and Jason Anderson and Zhuo Zhen and Pierre Riteau and Paul Ruth and Dan Stanzione and Mert Cevik and Jacob Colleran and Haryadi S. Gunawi and Cody Hammock and Joe Mambretti and Alexander Barnes and Fran\c{c}ois Halbach and Alex Rocha and Joe Stubbs},
  booktitle={Proceedings of the 2020 USENIX Annual Technical Conference (USENIX ATC '20)},
  publisher={USENIX Association},
  month={July},
  year={2020}
}

@article{lertpongrujikorn2025edgeweaver,
  title={EdgeWeaver: Accelerating IoT Application Development Across Edge-Cloud Continuum},
  author={Lertpongrujikorn, Pawissanutt and Kwon, Juahn and Nguyen, Hai Duc and Amini\_Salehi, Mohsen},
  year={2025},
  publisher={IEEE}
}

@inproceedings{paul2026oaas,
  author    = {Lertpongrujikorn, Pawissanutt and Amini Salehi, Mohsen and Paul, Tapajit Chandra},
  title     = {Object-as-a-Service (OaaS): Streamlining Cloud-Native Application Development for Edge-Cloud Continuum},
  booktitle = {Proceedings of the IEEE International Parallel and Distributed Processing Symposium (IPDPS) Tutorials},
  year      = {2026},
  month     = {May},
  note      = {Accepted}
}

@inproceedings{chanikaphon2023ums,
  title={Ums: Live migration of containerized services across autonomous computing systems},
  author={Chanikaphon, Thanawat and Salehi, Mohsen Amini},
  booktitle={GLOBECOM 2023-2023 IEEE Global Communications Conference},
  pages={467--472},
  year={2023},
  organization={IEEE}
}

@article{denninnart2023efficiency,
  title={Efficiency in the serverless cloud paradigm: A survey on the reusing and approximation aspects},
  author={Denninnart, Chavit and Chanikaphon, Thanawat and Amini Salehi, Mohsen},
  journal={Software: Practice and Experience},
  volume={53},
  number={10},
  pages={1853--1886},
  year={2023},
  publisher={Wiley Online Library}
}
